\RequirePackage{fix-cm}
\documentclass[natbib,smallcondensed]{svjour3}       % onecolumn (second format) for kais
\smartqed  % flush right qed marks, e.g. at end of proof
\usepackage[margin=2cm]{geometry}
\usepackage{graphicx,subfigure}
\usepackage{epstopdf}
\usepackage{verbatim}
\usepackage{graphicx}
\usepackage{caption}
\usepackage{mathtools}
\usepackage{dsfont}
\usepackage{multirow}
\usepackage{color}
\begin{document}

\title{CAPS: Context Aware Personalized POI Sequence Recommender System%\thanks{Grants or other notes
}
%\subtitle{}

%\author{}
%\institute{}

%\begin{comment}
\author{Ramesh Baral and Tao Li and XiaoLong Zhu}
\institute{Ramesh Baral \at
	School of Computing and Information Sciences, Florida International University,	Miami, FL, USA\\
	\email{rbara012@fiu.edu}           %  \\
	\and
	Tao Li \at
	School of Computing and Information Sciences, Florida International University,	Miami, FL, USA\\
	\email{taoli@cs.fiu.edu} %\\--------------------\\
	\and
	XiaoLong Zhu \at
	School of Computing and Information Sciences, Florida International University,	Miami, FL, USA\\
	\email{xzhu009@fiu.edu }
}
%\end{comment}
\date{}
\maketitle

\begin{abstract}
	\label{abstract}
The revolution of World Wide Web (WWW) and smart-phone technologies have been the key-factor behind remarkable success of social networks. With the ease of availability of check-in data, the location-based social networks (LBSN) (e.g., Facebook\footnote{www.facebook.com}, etc.) have been heavily explored in the past decade for Point-of-Interest (POI) recommendation. Though many POI recommenders have been defined, most of them have focused on recommending a single location or an arbitrary list that is not contextually coherent.
It has been cumbersome to rely on such systems when one needs a contextually coherent list of locations, that can be used for various day-to-day activities, for e.g., itinerary planning. 

This paper proposes a model termed as \textbf{CAPS} (\underline{\textbf{C}}ontext \underline{\textbf{A}}ware Personalized \underline{\textbf{P}}OI \underline{\textbf{S}}equence Recommender System) that generates contextually coherent POI sequences relevant to user preferences.
To the best of our knowledge, \textbf{CAPS} is the first attempt to formulate the contextual POI sequence modeling by extending Recurrent Neural Network (RNN) and its variants. CAPS extends RNN by incorporating multiple contexts to the hidden layer and by incorporating global context (sequence features) to the hidden layers and output layer. 
%This extension also helps the model to cope with the vanishing gradient problem. 
It extends the variants of RNN (e.g., Long-short term memory (LSTM)) by 
incorporating multiple contexts and global features in the gate update relations. 
%It exploits historical check-in data and personalizes the recommendation to generate contextually coherent sequences of places (equivalent to trip).
The major contributions of this paper are:
(i) it models the contextual POI sequence problem by incorporating personalized user preferences through multiple constraints (e.g., categorical, social, temporal, etc.),
(ii) it extends RNN to incorporate the contexts of individual item and that of whole sequence. It also extends the gated functionality of variants of RNN to incorporate the multiple contexts, and 
(iii) it evaluates the proposed models against two real-world data sets.
\end{abstract}

\keywords{Information Retrieval, Context Aware Recommendation, POI Recommendation, Social Networks}

\section{INTRODUCTION}
\label{sec:intro}
\label{sec:intro}
The POI recommendation systems have been very popular in the last few years. 
Such systems exploit the explicit historical check-in data and some implicit aspects or contexts to generate a list of POIs that are of potential interest to users. Generally, most of the POI recommenders recommend a single POI or an arbitrary list of POIs (\cite{yuan2013time, zhang2015geosoca}) that satisfy the personalized user preferences but may not be contextually coherent.
Often, users need a list of items that are contextually coherent as per the preferences of users. Such a coherent list of recommendation can be formulated as a sequence modeling problem where the sequence of items need to adhere to users' preferences by addressing constraints that are applicable to the consecutive items and also to the whole sequence. For instance, in the case of itinerary recommendation, the distance between consecutive places, total trip time, tentative start/end time, start/end place/category, etc. can be the major concerns. 

%The personalized POI sequence recommendation is more challenging than a single POI recommendation due to following major reasons : (i) it needs to find a set of POIs that satisfy the users' preferences, (ii) unlike the basket-item prediction in general sequence mining problems, the POIs in the recommended set should be coherent as they should satisfy multiple constraints, such as distance between the consecutive POIs, the popularity time of the POIs (recommended POIs should be relevant by time of day, for instance places popular for breakfast should be recommended in the morning time), the time budget constraint of the traveler, the social constraints (as the sequence of POIs are most likely a trip or tour where people prefer to travel in a group), categorical constraints (as the places with similar category can be an option), start and end POIs, visit durations, and (iii) the influence of the POIs visited earlier decay as time passes on.

%Travelers may prefer more specific and comprehensive recommendations, which take into account both user preferences and additional attributes, including travel durations and locations. It may be more helpful if a travel recommendation system can provide a travel plan, which contains a sequence of locations along with feasible travel routes.
In comparison to the general item recommendation, the POI recommendation is more challenging as it is sensitive to various aspects (e.g., categorical, social, spatial, temporal, etc.) (\cite{yuan2013time, zhang2015geosoca, baral2016maps, baral2016geotecs}) and the impact of those aspects might vary on the preference of user (~\cite{baral2017exploiting}). The POI sequence recommendation becomes even more interesting and challenging because of following major reasons:
\begin{enumerate}
	%\item an ideal POI sequence model needs to address the temporal popularity and temporal preferences because the POI visits during the peak time may result in a long wait time, poor service and sometimes a higher price,
	
	\item the brute-force approaches on computing all permutations of the POIs are NP-hard problem and are not useful,
	
	\item getting a preferred list from an arbitrary list may not be optimal because the preferred list need to satisfy multiple constraints, for instance, the places in a trip should be contextually coherent, need to satisfy many constraints, such as spatial (near/far places), temporal (relevant places for a time of a day, for instance bars can be relevant at evening or night), social (as the trip might be a family, friend focused), time budget constraint of the traveler, start and end POIs, etc.,
	
	\item the contexts can be user dependent and can vary dynamically in real-time (e.g., the same context may not be always relevant to a user).
	
	%\item unlike the language modeling tasks (where there is mostly a one-to-one mapping between the input and output items, e.g., language translation problems), the POI sequence modeling can have many-to-many relation (e.g., depending on the context, a user who has visited a set of locations can select multiple destinations). Also, the POI sequences can repeat (for instance, a user can depict same check-in pattern after some time interval) which is very rare in language modeling (e.g., in next sentence prediction, the input sentences are mostly unique, etc.). This makes the POI sequence prediction more challenging than other sequence modeling problems.

\end{enumerate}

There are few studies that focused on POI sequence recommendation. 
Most of the existing systems are focused on either frequency-based user interest or the average visit durations but incorporation of major constraints (such as distance between consecutive POIs, social constraints, categorical constraints, and temporal constraints) is less explored and is of great interest in the recommendation community.
%Although few of the existing studies have claimed the performance gain by exploiting different contextual attributes, the exploitation of contextual POI sequence modeling using Neural Networks has been barely explored. 
This paper attempts to fill the gap by incorporating multiple constraints for POI sequence modeling.

We can envision some implicit relevance between POI sequence modeling and the word sequence modeling. 
First, the smallest elements (check-ins and words) have some contextual coherence with their neighbors. Second, most of the subsequences (check-ins made within a time duration, and a sentence) comprise of contextually coherent elements and also have contextual relation with neighbors (check-ins of preceding/succeeding time duration, and neighbor sentences). Third, the ordered sequence of elements (chronological check-ins of a user, and a text) follow some pattern which can be modeled to learn relation between subsequences (e.g., predict the next check-in and predict next word).

%The words in language model can be related to the check-in records, the sentences can resemble an ordered subset of check-ins, and the text can resemble whole sequence of ordered check-ins for a user.
%The conceptual relevance between the two problem domains has envisioned the viability of exchange of state-of-art techniques across those domains. 
%This has inspired us to exploit the potential of techniques from language modeling domain and extend them for POI sequence recommendation.
Besides the aforementioned relevances, the contextual POI sequence modeling is more challenging due to the following reasons: (i) the POI sequence modeling can have longer contextual impact (e.g., unlike the language model where the n-grams can capture most of the context, the set of check-ins can be unique every day) and adhere to personalized preferences, 
%(ii) unlike language modeling, there are many-to-many relations between the context and target outcomes (for instance, at a given time of a day, a user can have multiple options to visit), 
(ii) unlike the language modeling tasks (where there is mostly a one-to-one mapping between the input and output items, e.g., language translation problems), the POI sequence modeling can have many-to-many relation (e.g., depending on the context, a user who has visited a set of locations can select multiple destinations). Sometimes the POI sequences can repeat (for instance, a user can depict same check-in pattern after some time interval) which is rare in language modeling (e.g., in next sentence prediction, the input sentences are mostly unique), and (iii) the real-time contexts can complicate the problem formulation using traditional sequence models.

Inspired from the wider popularity of RNN and its variants in sequence modeling (e.g., language modeling (machine translation) and image caption generation) (\cite{mikolov2010recurrent, mikolov2012context, ghosh2016contextual, bengio2003neural}), 
we attempt to cope with the challenges of contextual POI sequence modeling 
by incorporating the local contexts (context valid for subsequence) and global contexts (context valid for whole sequence). The local contexts (known as context now onwards in the paper) are incorporated into the recurrent module and the global contexts (known as feature of sequence now onwards in the paper) are fed to all the layers in the network. 
We also extend the gating mechanism of LSTM to incorporate the context and feature for personalized POI sequence modeling.
To the best of our knowledge, the proposed model is the first one to address the multi-context POI sequence modeling using extended RNN and its variants.

The major contributions of this paper are:
(i) it introduces all the major personalized user constraints (such as temporal, categorical, spatial etc.) as the influential context in the POI sequence modeling,
(ii) it extends RNN and its variants to explicitly incorporate relevant contexts,
%(iii) it incorporates the time-based preference decay for POIs in the recommendation model,
and (iii) it evaluates the proposed models against two real-world datasets.

\begin{figure}[h!]
	\centering     %%% not \center
	\subfigure[Check-in distribution in Weeplace dataset]{\label{fig:checkin_dist_weeplace}\includegraphics[width=0.45\textwidth,height=0.20\textheight]{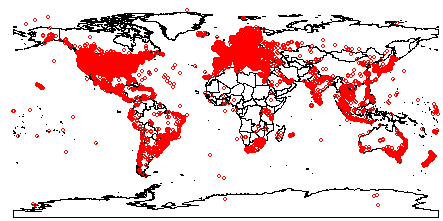}}
	\subfigure[Check-in distribution in Gowalla dataset]{\label{fig:checkin_dist_gowalla}\includegraphics[width=0.45\textwidth,height=0.20\textheight]{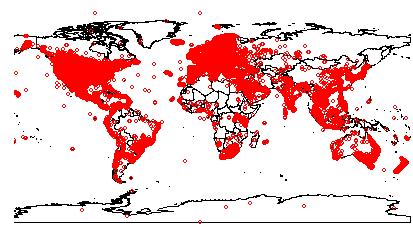}}
	\caption{Check-in distribution in different dataset}
	\label{fig:checkin_dist}
	\vspace{-1.5em}
\end{figure}

\begin{comment} 
(i) instead of using frequency-based user interest (by POI visit frequency) or requiring users to explicitly indicate their interest preferences,
we derive a relative measure of time-based user interest using a user's visit durations at POIs of a specific category, relative to the average visit durations of other users; 
(ii) we recommend a personalized POI visit duration to individual users based on their time-based user interests, instead of using the average POI visit duration for all users or not considering visit duration at all, 
(iii) we exploit different variants of Recurrent Neural Network (adapting context to the input, hidden, and output layers) to model the contextual recommendation,
(iv) we exploit multiple factors (e.g., temporal popularity, temporal user preferences, categorical constraint, spatial constraint - distance between consecutive pois, and social constraint - if the tour is feasible for group visit) while deriving the poi sequence.
\end{comment}

\section{RELATED RESEARCH}
\label{sec:rel_res}
We categorize the relevant studies using two broad notions - general data mining-based studies, and neural network-based studies.
\subsection{General Data Mining Approaches}
Most of the existing studies were focused on analysis of landmark visit patterns by exploiting the geo-spatial and temporal traces from check-ins. 
However, those studies mostly focused on the analysis only and ignored synthesis or recommendation of new paths.
The Orienteering approach from~\cite{feillet2005traveling} (a variant of Traveling salesman problem) was the dominating technique for earlier studies.
This approach was focused on the objective of finding a subset of places that could maximize the reward under some constraints, such as maximum travel cost (e.g., the time limit is not exceeded). 
The other widely exploited techniques were collaborative filtering (\cite{zhang2015personalized, yu2016personalized}), content-based approach (\cite{bohnert2012geckommender}), Apriori principle (\cite{lu2012personalized, yu2016personalized}), matrix factorization (\cite{ge2011cost}), topic-modeling (\cite{liu2011personalized, liu2014cocktail}), and tree traversal approaches (\cite{zhang2015personalized}).
\cite{dunstall2003automated} proposed a trip planner that matched the explicit user inputs against a database to get the best matching itineraries. 
%together structured components of types “tour”, “lodging”, and “transportation”. Tours typically contain places to visit and activities to perform within a single day and general area, i.e. between lodgings and transports. 
\cite{tai2008recommending} mined itineraries' data and generated sequences by implicitly incorporating users' interests via the images on Flickr\footnote{https://www.flickr.com}. However, they did not focus on constructing an itinerary from scratch and also did not address the major aspects, such as the temporal influence on users and locations.

A content-based filtering approach from \cite{bohnert2012geckommender} 
exploited visitors' explicit ratings to recommend tours in a museum.
%The predicted ratings in turn form the basis for theme/tour recommendations. 
Unlike a real trip, the order of recommended items and the distance between consecutive items were not focused in this system.
The greedy algorithm-based approaches (\cite{choudhury2010automatic, vansteenwegen2011city}) matched user profiles (consisting of preferred location categories, types, start/end time, and keywords) to relevant places.
%Choudhury et al.~\cite{choudhury2010automatic} extended the Orienteering problem and exploited geo-tagged Flickr photos to infer user interest and recommended multi-day itineraries using a recursive greedy algorithm.
\cite{lu2012personalized} mined user check-in data and exploited the Apriori-based algorithm. Thought it was computationally intensive, it was able to find optimal trips under multiple constraints, such as social relationship, temporal property, categorical diversity, and parallel computing simultaneously.
%It was the first work on travel recommendation that adapted multiple constraints, such as social relationship, temporal property, categorical diversity, and parallel computing simultaneously. They first evaluated the personalized score of each attractions by considering the user preferences and temporal constraints. The top-k attractions were then subjected with multiple constraints. Two different scores (user-based and temporal-based) of locations were computed using collaborative filtering and temporal popularity of the locations. These linear fusion of these two scores were used during the trip planning.
%In ~\cite{gunawan2014mathematical}, the uncertain wait time problem was addressed using a meta-heuristic algorithm to plan tours for theme park visitors. In~\cite{gavalas2015heuristics}, a time-varying travel cost was modelled and trips were recommended for a group of users. These studies mainly focused on the trade-off between efficiency and optimality of the solution rather than the relevance of the trips to the users. Moreover, the trips are not personalized and the same origin destination pair will result in the same trip for all users.
\cite{zhang2015personalized} exploited the uncertain travel time between two POIs and the POI availability constraints and proposed a tree-based algorithm to solve the personalized trip recommendation problem.
A recent study from \cite{wang2016improving} proposed a personalized trip recommender that was able to handle the crowd constraints (i.e., avoid peak hour of POIs) by extending the Ant Colony Optimization algorithm. 
\cite{zheng2011learning} proposed a HITS-based inference model to evaluate the degrees of user experiences and location interests based on GPS trajectory data. 
%This study considered that interesting places might be accessed by many greater travel experts and experienced tourists would visit many interesting places. They further discovered the classical travel sequences among locations. However, GPS trajectory data is comparatively difficult to obtain and still not readily available.

A heuristic-based algorithm from \cite{chen2015tripplanner} exploited user preferences, and traveling times based on different traffic conditions using trajectory patterns derived from taxi GPS traces. The heuristic search-based travel route planning system from \cite{yu2016personalized} exploited only user preferences and spatio-temporal constraints. 
%They adapted the collaborative filtering algorithm to discover and rank the POIs; then, we define the transition likelihood from one location to another to model user and physical constraints. They also used apriori-based sequence generation to recommend travel plans. The computational cost is expensive in bigger dataset, social, diversity factors were not addressed in their study.
\cite{jiang2016personalized} proposed a ranking-based system to recommend personalized travel sequences in different seasons by merging textual data and view point information extracted from images. However, they did not use the constraints for social, temporal preference of users, and temporal popularity of places.
Another recent study from \cite{lim2017personalized} proposed time-based user interest and demonstrated advantages over the use of frequency-based popularity measures.% in tour personalization. 
They exploited geo-tagged images and incorporated contexts, such as visit duration, users' preferences, and start and end points. However, their model did not address the categorical, temporal, and social constraints on the POI sequence.
%They exploited the geo-tagged images to analyze the visiting pattern of users and recommended personalized tours using POI popularity and user interest preferences, which were automatically derived from real-life travel sequences based on geo-tagged photographs. 
A probabilistic model from \cite{chen2016learning} incorporated POI categories and user behavior history to generate trajectories. They used Rank-SVM to rank the items and used Markov model to predict the transition between POIs. 
%They incorporated category, popularity (number of distinct visitors), total number of visits and average visit duration for each POI.

%Most of the existing studies focused on the preference of individual user and have not explored the preferences of group of users (e.g., friends). 
Most of the existing studies have exploited few contexts and have focused on personalized POI visit durations. Unlike others, our study exploits multiple contexts (temporal, spatial, categorical, and social) to generate the POI sequence. 
%It also models the visit duration of a place by aggregating the visit duration of places with similar category.
In comparison to the existing studies, this paper presents following major differences:
(i) it exploits multiple contexts (temporal, spatial, categorical, and social) to model the contextual POI sequence problem,
(iii) it extends the RNN by incorporating multiple contexts to efficiently model the personalized contexts in POI sequence generation. It extends the gating mechanism of variants of RNN to incorporate multiple contexts for efficient POI sequence modeling, and
%(ii) it aggregates the visit durations of a place by the visit duration of places with similar category,
(iii) it evaluates the generated sequences with various relevant metrics (e.g., pairs-F1 score, diversity, displacement (see Section~\ref{ssec:eval_metrics} for detail)) on two real-world datasets.

\subsection{Neural Network-based approaches}

The Recurrent Neural Networks and its variants (e.g., Long Short-Term Memory (LSTM), Gated Recurrent Unit(GRU), etc.) were quite effective in sequence modeling problems (for instance speech recognition and machine translation) (\cite{mikolov2012context, twardowski2016modelling, sutskever2011generating}) that can exploit the dependencies between spatially correlated items (such as words in a sentence and pixels in images).
Though some of the neural network-based models focused on predicting trajectories, they were more focused on collision avoidance between humans (\cite{alahi2016social}). The Social LSTM from \cite{alahi2016social} extended the approach from ~\cite{graves2013generating} by incorporating the spatial distance between neighbor objects and predicted the trajectory for human movement.
\begin{comment} the source code is available at authors page at http://web.stanford.edu/~alahi/\end{comment}
The existing models were mainly focused on collision avoidance and ignored the constraints (such as user preferences and POI contexts) relevant to POI sequence generation.

Though RNNs were quite popular in sequence modeling for image and text, relatively fewer studies have exploited it for the recommendation domain. A recent study from \cite{wu2017recurrent} used the Recurrent Recommender Networks (RRN) to predict future behavioral trajectories. They exploited LSTM (\cite{hochreiter1997long}) based model to capture the user-movie preference dynamics, in addition to a more traditional low-rank factorization.
%This was achieved by endowing both users and movies with a Long Short-Term Memory (LSTM)~\cite{hochreiter1997long} autoregressive model that captured preference dynamics, in addition to a more traditional low-rank factorization. They used a nonparametric model that is able to extrapolate behavior into the future by learning the inherent user and movie dynamics. Their study was more focused on predicting the rating of items given the previous ratings and use the latest prediction to get the next best rating prediction.
\cite{hidasi2015session} trained a GRU (\cite{cho2014learning})-based model with a ranking loss to predict the next item in the user session on e-commerce system. However, they did not consider personalization. 
A contextual network from \cite{smirnova2017contextual} used three different techniques (concatenation, multiplication, and both) to integrate the context embedding with the input embedding and was used for next item prediction.
Though some of the neural network-based recommenders exist, none of them focused on multi-context personalized POI sequence modeling.

\section{Methodology}

In this section, we define our proposed approach. First, we define the basic concepts that are frequently used in the paper and then we elaborate the incorporation of contexts into RNN and its variants.

\paragraph{\textbf{Context}}: A context of a check-in represents the current and previous scenarios which have direct or indirect influence on the selection of next POI. Such a context (for instance current time, category of current place, category of previous place, popularity score of current place, and so forth.) can be represented as a high-dimensional vector.

\paragraph{\textbf{Context-aware POI sequence}}: Given a set of contexts C = \textbraceleft ${c_i}$\textbraceright,  where $i \in K$, is a context type, our objective is to predict or generate a sequence of POIs, that are most relevant to the given context and match the user preferences.

Given a user \textit{u} who has made n check-ins, we define her travel history as an ordered sequence, $H_u = (V_1, V_2,..., V_n)$, where $V_i$ is a check-in activity and can be represented by the triplet $V_i = (l_i,a_i,d_i)$ indicating the location ($l_i$) of the check-in, arrival time ($a_i$) to $l_i$, and the departure time ($d_i$) from $l_i$ respectively. 
%\textcolor{red}{Make sure the dataset has the start and end time of a check-in. If we donot have the end time of the check-in, aggregate this with the minimum of the difference of the start time to this location and start time to other location, and the travel time (also aggregate with the distance between this location and the next location)}.
The travel sequences can be split into small subsequences (for instance, the check-ins made within a time interval, e.g., 8 hours, one day, etc.). As in earlier studies (\cite{lim2015personalized, choudhury2010automatic}), we use the time interval of 8 hours.
\paragraph{\textbf{Visit duration of POI}}:
The average visit duration (stay time (ST)) of a POI \textit{i} is defined as:
\begin{equation}
\label{eqn:time_spent}
ST(i) = \frac{1}{\mid U \mid} \sum\limits_{\substack{u\in U \\ V_u \in H_u \\V_u.l = i}} \frac{1}{\mid V_{u,i} \mid}\sum\limits_{l \in V_{u,i} } (a_{l+1} - TT(l, l+1) -a_l ),
\end{equation}
where U is the set of all users, $V_{u,l}$ is the set of visits made by the user \textit{u} to location \textit{l}, TT(a,b) is the travel time between POI \textit{a} and POI \textit{b}. 
We can use a log normal distribution to compute the travel time between two POIs visited consecutively. The stay time is 0-1 normalized and is represented as $ST'(i)$.
%The travel time can be calculated using the distance between the two locations and taking an average travel time (~30 mph) (or in an extreme case we can use other heuristics like the speed limit of the city)
%Various studies and methods have been proposed to estimate travel time distributions in the literature. See ~\cite{guessous2014estimating} for a comparison of these methods. Zhang et al.~\cite{zhang2015personalized} used log normal distribution for this.
%\begin{equation*}
%TT(a, b) = dist(a, b)/30
%\end{equation*}

We define the user interest to a place in terms of aggregate of stay time (AST) to that place. This term in turn relies on the visit frequency and the stay time to that place, and the stay time to the places of same category.
\begin{align}
\label{eqn:aggregate_time_spend}
AST(u, i)_{cat} = (1-\alpha) * \mathcal{F}_{st'}(u,i) + \alpha * \mathcal{G}_{st'}(u,i) \sum\limits_{\mathclap{\substack{l \in V_u \\ l.cat=i.cat}}} \frac{ST'(l)}{V'_{u,l}},
\notag \\
\mathcal{F}_{st'}(u,i) = \frac{ST'(i)}{V'_{u,i}} \text{ if $\mid V_{u,i} \mid>0$, i.e. \textit{u} has some visits to location \textit{i}}, \notag \\
\mathcal{F}_{st'}(u,i) = 0 \text{ otherwise,} \notag \\
\mathcal{G}_{st'}(u,i) = \frac{1}{\sum\limits_{\mathclap{\substack{l \in V_{u} \\ l.cat=i.cat}}} 1} \text{ if $\exists V_{u,l} \wedge l.cat =i.cat $, i.e., \textit{u} has some check-ins on category \textit{i.cat} ,} \notag \\
\mathcal{G}_{st'}(u,i) = 0 \text{ otherwise,}
\end{align}
where $V_u$ is the set of visits by user \textit{u}, $V_{u,l}$ is the set of visits by user \textit{u} to location \textit{l}, $V'_{u,l} = \frac{\mid V_{u,l}\mid}{\mid V_{u} \mid}$ is the normalized visit count of user \textit{u} to location \textit{l}, $l.cat$ is the category of location \textit{l}, and $\alpha$ is a constant tuning factor estimated using the fraction of check-ins that are of same category as location $i$ .
After incorporating social impact, the average stay time on a location \textit{i} can be defined as:
\begin{align}
\label{eqn:aggregate_time_spend_friend}
AST(u, i) = (1-\psi_1) * AST(u, i)_{cat} + \psi_ * \mathcal{J}(F_{u})\sum\limits_{\mathclap{k\in F_u}}AST(k, i)_{cat} \notag, \\
 \mathcal{J}(F_{u}) = \frac{1}{\mid F_u\mid} \text{ if $\mid F_u\mid >0$ , i.e. \textit{u} has some friends,} \notag \\
 \mathcal{J}(F_{u}) = 0 \text{ otherwise,}
\end{align}
where $F_u$ represents the set of friends of user $u$, $\psi_1$ is a constant tuning parameter estimated using the fraction of check-ins of the user \textit{u} that are common to her friends.
Similarly, the average stay time by a user \textit{u} to a location category \textit{'cat'} can be defined as:
\begin{equation}
\label{eqn:aggregate_time_spend_friend_cat}
AST(u)_{cat} = (1-\gamma_1) * (\sum\limits_{\mathclap{\substack{i \in V_u \\ i.cat = cat}}} AST(u,i)_{cat}) + \gamma_1 *(\sum\limits_{\mathclap{\substack{j \in F_u \\k \in V_j \\ k.cat = cat}}} AST(j,k)_{cat}),
\end{equation}
where $\gamma_1$ is a tuning factor estimated using the fraction of check-ins of user \textit{u} that are common to her friends and have category \textit{'cat'}. $AST_{cat}$ is the aggregate of average stay on the category \textit{'cat'} from all users and $AST^{t}_{cat}$ gives the measure for time \textit{t}.
\paragraph{\textbf{Preference score of POI}}:
For a user \textit{u}, the preference score (PS) of a place \textit{l} at time \textit{t} is defined using the following relation:

\begin{align}
\label{eqn:preference_score}
PS(u,l, t) = \beta * \text{\textbraceleft} (1-\theta) * \mathcal{P}(u,l)*\mid V_{u,l,t} \mid + \theta * \mathcal{Q}(u,l)\sum\limits_{\mathclap{\substack{l' \in L\\ l'.cat = l.cat}}} \frac{ \mid V_{u,l',t} \mid }{\mid V_{u,l'} \mid} \text{\textbraceright}
 + (1-\beta)*AST(u, l) \notag, \\
\mathcal{P}(u,l) = \frac{1}{\mid V_{u,l} \mid} \text{ if $\mid V_{u,l} \mid >0$} \notag, \\
\mathcal{P}(u,l) = 0 \text{ otherwise} \notag, \\
\mathcal{Q}(u,l) = \frac{1}{\sum\limits_{\mathclap{\substack{l \in V_{u} \\ l.cat=i.cat}}} 1} \text{ if $\exists V_{u,l} \wedge l.cat =i.cat $, i.e., \textit{u} has some check-ins on category \textit{i.cat} } \notag, \\
\mathcal{Q}(u,l) = 0 \text{ otherwise,}
\end{align}
where $V_{u,l,t}$ is the set of visits made by user \textit{u} to location \textit{l} at time \textit{t}, L is the set of all locations, $\theta$ can be estimated as in Equation~\ref{eqn:aggregate_time_spend}, and $\beta$ is a tuning factor estimated using TF-IDF (term frequency inverse document frequency) (\cite{salton1986introduction}).
This relation addresses the trade-offs between visit frequency and stay time which is crucial to reward the preferred check-ins with low frequency but reasonable stay time.
The generalized preference score $PS(l,t)$ can be easily derived from above relation by considering the visit frequencies of all the users and stay time of all the users to this location at time \textit{t}.
Furthermore, the usage of categorical and the social contribution can handle the cold-start items (items with no check-ins information) and cold-start users (users with no check-ins) to some extent as the relations defined above incorporate some contributions from categorical and social factors. 
We also handle the scenario in which places with higher preference scores are compromised due to other constraints, such as travel time and cost. 
To address the trade-off between constraints and preference score, we define a consolidated preference score as:
\begin{equation}
\label{eqn:cons_score}
P(u,l,t) = PS(u,l,t)*(1 - \frac{1}{m}\sum\limits_{\mathclap{i=1}}^{m}Constraint_{i}(l,p)),
\end{equation}
where $Constraint_{i}(l,p)$ is a normalized numeric measure of $i^{th}$ constraint between the users' current location $p$ and the target location $l$.
For instance, the spatial constraint is the measure of distance between locations \textit{p} and \textit{l} which is min-max normalized by the minimum and maximum distance traveled by any user to reach location \textit{l} from any other location.
The above mentioned preference metrics are used as features and attributes (defined later) when we train our prediction model.
\subsection{Network Design}
We exploit RNNs because they are ideal to our problem due to the following reasons:
(i) the POI check-in patterns are relatively casual and vary across the users. Thus a long term dependency is required which can be modeled by RNNs (or its variants like GRU and LSTM),
(ii) the check-in trends are associated with many contextual attributes, such as social, categorical, temporal, and spatial. The capability of RNNs to incorporate additional attributes (e.g.,~\cite{mikolov2010recurrent, mikolov2012context}) is another reason behind selecting them.
%Our problem can be formulated as the language modeling domain by just replacing the sequence of words with the sequence of check-ins along with the additional contextual parameters. 
%These additional parameters are fed in the form of additional nodes which represent the spatial, temporal, categorical, and social constraints of the sequence.

\begin{figure}[h!]
	\centering     %%% not \center
	\subfigure[POI Sequence Modeling with basic RNN]{\includegraphics[width=0.45\textwidth,height=0.20\textheight]{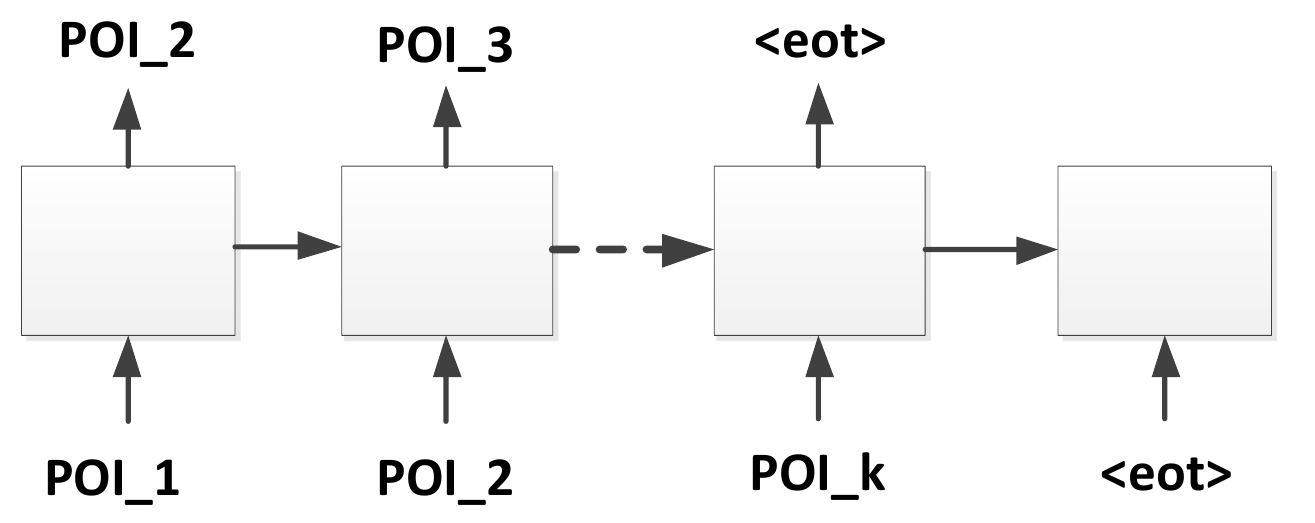}\label{fig:arch_a}}
	\subfigure[Contextual POI Sequence Modeling]{\includegraphics[width=0.45\textwidth,height=0.20\textheight]{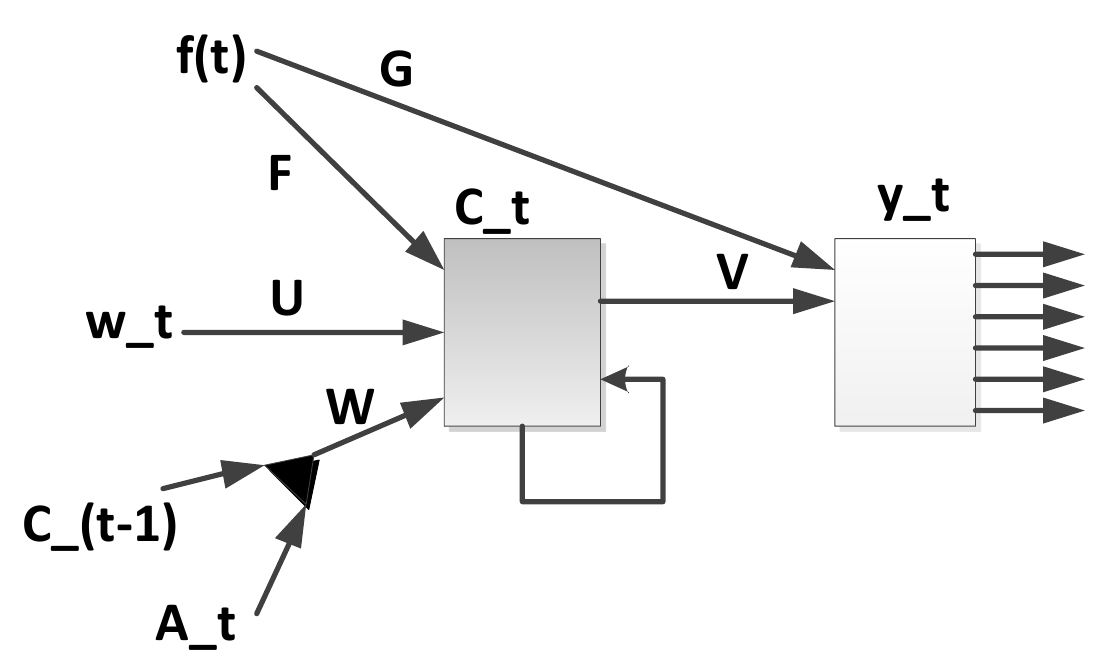}\label{fig:arch_b}}
	
	\caption{Overview of proposed system architecture. Figure a shows the sequence modeling with general RNN where no contextual information is used in the hidden and output layers. Figure b shows the sequence modeling with extended RNN where the contextual information is explicitly fed to the  hidden and output layer.}
	\label{fig:arch_combined}
	\vspace{-1.5em}
\end{figure}

\begin{figure}[h!]
	\centering
	\includegraphics[width=0.7\columnwidth,height=0.35\columnwidth]{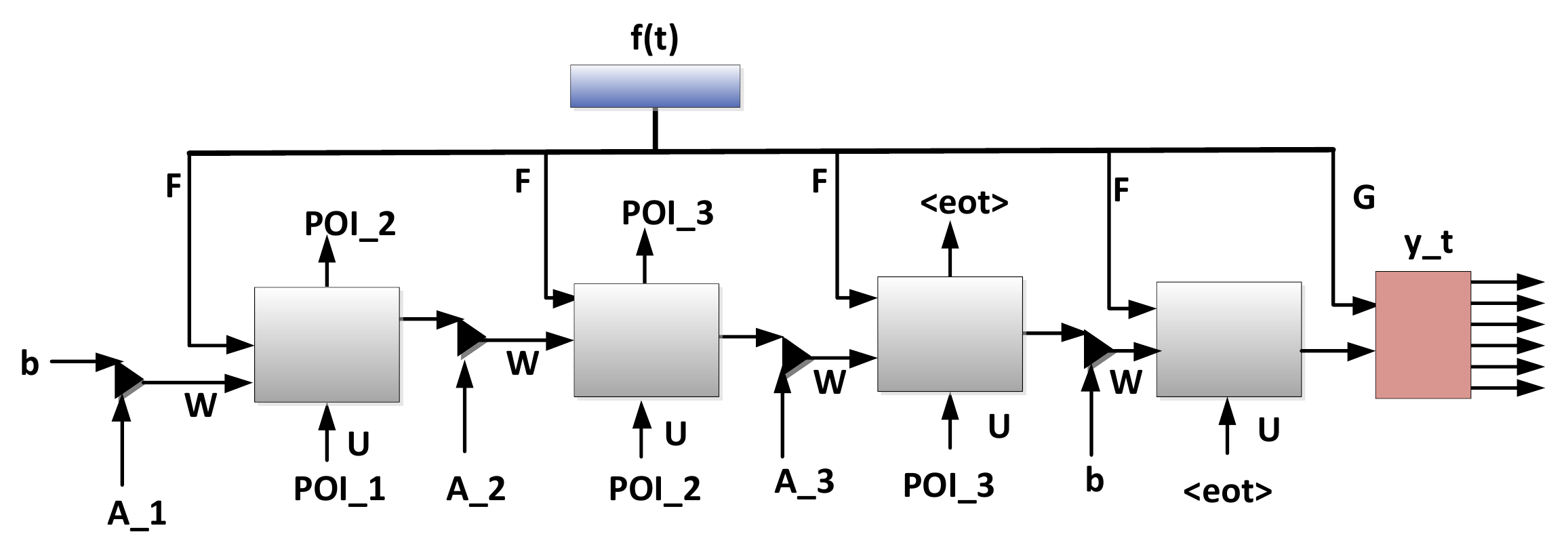}%
	\caption{Unraveled view of Contextual POI Sequence Model representing the propagation of contextual information through different layers of the network.}%
	\label{fig:arch_c}
	\vspace{-1.5em}
\end{figure}
\subsubsection{Preliminaries of proposed Network}
RNNs are ideal for sequence modeling problem because they can efficiently model the generative process of sequential information, can summarize the information in hidden states%(a continuous representation)
, and can generate new sequences by using the probability distribution specified by the hidden states. The hidden state $h_t$ from the input sequence ($x_1, x_2, ...., x_t$) is recursively updated as $h_t = f(h_{t-1}, x_t)$, where $f(.,.)$ is a non-linear transformation function (for e.g., $h_t = tanh(Uh_{t-1} + Vx_t) $, where U and V are weight matrices). The overall probability of a sequence $ \overrightarrow{x} = x_1, x_2,.....,x_N$ can be defined as:
\begin{equation}
p(\overrightarrow{x}) = \prod_{t=1}^{N} p(x_t\mid h_{t-1}).
\end{equation}
%The above equation can be easily adjusted to incorporate the context in the hidden state.
Training such a regular RNN by maximizing the above function suffers from vanishing gradient (i.e., a problem with gradient based learning methods where the gradient (error signal) decreases exponentially with the value of n (number of layers) while the front layers train very slowly, see~\cite{hochreiter1998vanishing} for further detail). While the architectural enhancements, such as long short-term memory (LSTM) from~\cite{hochreiter1997long} and Gated Recurrent Unit (GRU) from~\cite{cho2014learning} are believed to address the vanishing gradient problem, the context enriched models (\cite{mikolov2010recurrent, mikolov2012context}) are also claimed as an effective solution. We formulate the contextual POI sequence modeling by incorporating the contexts into RNN and its variants. 

%The architecture of the proposed model is illustrated in Figure~\ref{fig:arch_combined}. 
The basic POI sequence modeling with regular RNN is illustrated in Figure~\ref{fig:arch_a}. The proposed contextual POI sequence modeling is illustrated in Figure~\ref{fig:arch_b}. The unraveled view of the contextual RNN (Figure~\ref{fig:arch_b}) is illustrated in Figure~\ref{fig:arch_c}.
The input vector and the output vector both have the dimension of the number of POIs. The input vector $w(t)$ is the check-in at time t (which is a one hot encoding), and the output layer produces a probability distribution over the POIs, given the context, feature and the previous check-in. The hidden layers of RNNs are responsible to maintain the history of check-ins that form the sequence. Inspired from \cite{mikolov2012context}, we extend the vanilla RNN to incorporate the context and feature inputs (f(t)). The context and the attribute of the current input, the current input itself, and the features representing the whole sequence are fed to the network. The features are propagated to the hidden as well as to the output layers. This helps us to retain the important features which the network will be aware at each and every iteration (see Figure~\ref{fig:arch_b}).

\subsubsection{Basic POI Sequence Modeling}
In this model, we use the regular RNN which is popular for modeling the sequential data. The capability of maintaining the session information into the recurrent states are often considered as its strength. Although they are theoretically claimed as poor candidates for problem with long session relations (due to vanishing gradient problem), architectural extensions (e.g., LSTM) or contextual extensions (e.g., ~\cite{mikolov2012context}) are some of the state-of-art techniques in language modeling domain.

With respect to POI sequence modeling, the RNNs can use the probability distribution specified by the hidden state to predict the next POI. The hidden state summarizes the information of the sequence observed so far (e.g., $l_1, l_2,....,l_{t-1}$).
A hidden layer depends on the current input ($x_t$) and the input at previous step ($h_{t-1}$), i.e. $h_t = f(W_{xh}x_t + W_{hh}h_{t-1})$, where $W_{xh} \text{ and } W_{hh}$ are the weights for the current input, and for the value from previous hidden layer respectively.
The probability of a POI sequence can then be defined using the chain rule:
\begin{equation}
\label{eqn:prob_seq}
p(\overrightarrow{l}) = \prod_{t=1}^{T}p(l_t \mid h_{t-1}),
\end{equation}
where T is the number of timesteps to be considered.
The negative log likelihood is used as the sequence loss to train the network:
\begin{equation}
\mathcal{L}(x) = - \sum\limits_{t=1}^{T}log(p(l_t \mid h_{t-1})).
\end{equation}

For the defined network, the partial derivative of the loss can be computed using the backpropagation through time (\cite{williams1995gradient}) and the network can be trained using gradient descent.
Such a sequence modeling is illustrated in Figure~\ref{fig:arch_a}. 
%In this case, the network learns to predict the next POI by using the input data that contains previously seen POI sequences. 
This network predicts the next POI by learning from the previously fed POI sequences only and does not consider other information, such as POI attributes, current context, and other relevant features. The contextual model defined in next subsection addresses these aspects.

\subsubsection{Contextual POI Sequence Modeling using RNN}
This model is termed as \textbf{CAPS-RNN} and represents the contextual POI sequence modeling as illustrated in Figure~\ref{fig:arch_b} and Figure~\ref{fig:arch_c}. 
CAPS-RNN extends the regular RNN by incorporating the context to hidden layers and the feature (f(t)) to the hidden and output layers. This extension has following major advantages:
(i) as the regular RNNs have no provision of explicit context propagation and as the POI sequence modeling requires the contexts to be remembered for longer span, the extension helps us to propagate the relevant contexts to make sure that the next predicted subsequence adheres to the context,
(ii) the regular RNNs suffer from vanishing gradient problem (i.e. during back-propagation, the gradients decrease exponentially with the value of n (number of layers) and the front layers train very slowly, see~\cite{hochreiter1998vanishing} for further detail). The explicit feeding of contexts to the hidden layer and the features to hidden and output layers helps alleviate the gradients, meanwhile retaining the relevant context (see \cite{mikolov2012context} for detail). 

The feature contains information that is relevant to the current input and also to the whole sequence, for instance with respect to POI sequence modeling, it can be the time budget constraint, start/end place categories, the geographical vicinity of interest, etc.
These are the auxiliary information that can have significant impact on the prediction tasks.
The context represents the aspect that is relevant to the current input (for instance, the current time, previous check-ins which impact the current check-in).
In addition to context and feature, the attribute vector which represents any information that is relevant to the current item (for e.g., the locations' category, locations' hours, locations' popularity, locations' preference score, and locations' average stay time) is also incorporated to the network. For the sake of ease, the attribute and context vectors can be combined together.
%We use the simple multiplication of the context vector and the attribute vector as shown in Figure~\ref{fig:arch_b}. 

After we train the network (using stochastic gradient descent), the output vector y(t) gives the probability measure of different POIs at time \textit{t}, given the previous POI, context, and the feature vector. The hidden and output layers are updated using the following relations:
\begin{equation}
c(t) = \mathcal{H}(\textbf{U}w(t) + \textbf{W}(c(t-1)\odot A(t)) + \textbf{F}f(t)),
\end{equation}
\begin{equation}
\hat{y}_t = \mathbf{V}c(t) + \mathbf{G}f(t),
\end{equation}
\begin{equation}
y(t) = g(\hat{y}_t),
\end{equation}
where $\mathcal{H}(.)$ is a non-linear function (e.g., tanh), A(t) is the context attribute vector of current item for time \textit{t} (see Eqn.~\ref{eqn:attribute_vector}), f(t) is the sequence feature vector (see Eqn.~\ref{eqn:feature_vector}), and $g(a_i) = \frac{\exp({a_i})}{\sum\limits_{j}\exp({a_j})}$ is the softmax function.
The sequence feature vector contains the information relevant to the whole sequence. 
We use the feature vector of following format:
\begin{equation}
\label{eqn:feature_vector}
f(t) = <cat_{start}, cat_{end}, loc_{start}, loc_{end}, loc_{dist}, time_{start}, time_{end}>,
\end{equation}
where $cat_{start}$ is category of starting place, $cat_{end}$ is category of ending place, $loc_{start}$ is starting place, $loc_{end}$ is ending place, $loc_{dist}$ is the distance between consecutive places, $time_{start}$ is starting hour, and $time_{end}$ is ending hour of the POI sequence. All the non-numeric elements of this vector are label encoded before feeding to the network.
The attribute vector contains the information relevant to the current item and current context. We use the following format for the attribute vector:
\begin{equation}
\label{eqn:attribute_vector}
A(l, t) = <ST'(l), AST(l), AST^{t}_{cat}, PS(l), l.cat, l_T, l_{dist}>,
\end{equation}
where $l_T = l_1, l_2,...,l_T$ is the temporal popularity of location \textit{l} for each hour, \textit{'cat'}=$l.cat$ is category of location \textit{l}, and $l_{dist}$ is the distance of location $l$ from the previous location in the sequence. All the non-numeric elements of this vector are label encoded.
The feature vector and the attribute vector can be used to incorporate additional features and attributes if required.
The network is trained to learn the weight matrices (\textbf{U}, \textbf{V}, \textbf{W}, \textbf{F}, and \textbf{G}), and to maximize the likelihood of the training data (\cite{mikolov2012statistical, bengio2003neural}). The probability of a POI sequence \textit{l} for the network can be defined as:
\begin{equation}
p(l) = \prod_{t=1}^{T}p(l_{t+1}\mid y_t).
\end{equation}

\subsubsection{Contextual POI Sequence modeling using LSTM}

This model is termed as \textbf{CAPS-LSTM}. The core idea behind LSTM is the introduction of memory state and multiple gating functions to control the write, read, and removal (forget) of the information written on memory state. 
The information is propagated by applying these gates to the input data and data from the previous memory states.
\begin{comment}
The state update equations of LSTM are listed below:
\begin{align}
i_t = \sigma(W_ix_t + W_{hi} h_{t-1} + W_{ci}c_{t-1} + b_i) \notag \\
f_t = \sigma(W_fx_t + W_{hf} h_{t-1} + W_{cf}c_{t-1} + b_f) \notag \\
z_t = tanh(W_zx_t + W_{hc} h_{t-1} + b_z) \notag \\
c_t = f_t \odot c_{t-1} + i_t \odot z_t \notag \\
o_t = \sigma(W_ox_t + W_{ho} h_{t-1} + W_{co}c_{t-1} + b_o) \notag \\
h_t = o_t \odot tanh(c_t)
\end{align}
\end{comment}
We incorporate the explicit context to each LSTM cell because each cell models a subsequence and each subsequence can have potentially unique context. This explicit context to each LSTM cell and the feature of sequence to all LSTM cells help us propagate the relevant context for the sequence modeling.
Due to space constraint, we only provide the update equations of our extended LSTM which incorporates the contextual information. The hidden layer of contextual LSTMs are updated using the following relations:
\begin{align}
i_t = \sigma(\textbf{W}_ix_t + \textbf{W}_{hi} (h_{t-1}\odot A_t) + \textbf{W}_{ci}c_{t-1} + b_i + \textbf{W}_fF), \notag \\
f_t = \sigma(\textbf{W}_fx_t + \textbf{W}_{hf} (h_{t-1}\odot A_t) + \textbf{W}_{cf}c_{t-1} + b_f + \textbf{W}_fF), \notag \\
z_t = tanh(\textbf{W}_zx_t + \textbf{W}_{hc} (h_{t-1}\odot A_t) + b_z + \textbf{W}_fF), \notag \\
c_t = f_t \odot c_{t-1} + i_t \odot z_t, \notag \\
o_t = \sigma(\textbf{W}_ox_t + \textbf{W}_{ho} (h_{t-1}\odot A_t) + \textbf{W}_{co}c_{t-1} + b_o + \textbf{W}_fF), \notag \\
h_t = o_t \odot tanh(c_t).
\end{align}
where $c_t$ is the memory state, $z_t$ is the module that transforms information from input space $x_t$ to the memory space, and $h_t$ is the information read from the memory state. 
%The terms $i_t$, $f_t$, $o_t$ represent the input, forget, and output gates respectively. 
The input gate $i_t$ controls information from input $z_t$ to memory state, the forget gate $f_t$ controls information in the memory state to be forgotten, and the output gate $o_t$ controls information read from the memory state. The
memory state $c_t$ is updated through a linear combination of input filtered by the input gate and the previous memory state filtered by the forget gate. The term $\textbf{W}_f$ is feature weight matrix, $F$ is feature vector, $A_t$ is attribute vector as in the contextual RNN model, and $\odot$ is element-wise product operator. The relevant weight matrices \textbf{W} and biases $b$ are subscripted accordingly.

%\subsubsection{Decay function}
%\textcolor{red}{The recent visits have more impact than the visits in the past. This is addressed by the decay function....}

\subsubsection{Sequence generation}
%We train one network per user. 
After the model is trained, it can generate the personalized POI sequences. 
%The sequence generation has already been explored in language modeling domain~\cite{sutskever2011generating, graves2013generating}.
The basic idea is to train the network using a sequence data one step at a time. The sampled output from the network is then fed as the input for the next level. If we have k different POIs, and $k^{th}$ POI is checked-in at time $t$, then the input $x_t$ is one-hot encoded vector with only the $k^{th}$ entry set to 1. The output of the network is a multinomial distribution which is parameterized using a softmax function and can be defined as:
\begin{equation}
p(x_{t+1} = k \mid y_t) = y_{t}^{k} = \frac{\exp(\hat{y}_{t}^{k})}{\sum\limits_{k'=1}^{K} \exp(\hat{y}_{t}^{k'})}.
\end{equation}
From the generated sequences, the top-k scorers (sum of preference scores (AS) of all places in a sequence) are recommended to user.
\begin{comment}
\textcolor{red}{Section 3.1.5 could give some additional detail on the generation process and the evaluation procedure (maybe adding some examples?).}
\end{comment}
\section{Evaluation}
\label{sec:eval}
In this section, we describe the dataset, the evaluation baselines, evaluation metrics, and the experimental results.
\subsection{Dataset}
\label{ssec:dataset}
We used two real-world datasets collected from two popular LBSNs -Gowalla and Weeplaces (\cite{liu2013personalized}). These datasets were well defined and have the attributes relevant to the context of this paper, such as (i) the location category, (ii) geospatial co-ordinates, (iii) friendship information, and (iv) check-in time. The statistics of the dataset is summarized in Table~\ref{tab:dataset_stat}.
%The Weeplaces dataset has 7,658,368 check-ins from 15,799 users over 971,309 different locations. The Gowalla dataset has 36,001,959 check-ins from 319,063 users over 2,844,076 locations. 

\begin{table}[h!]
	\centering
	\begin{tabular}{|c|c|c|c|c|c|}
		
		\hline \rule[-2ex]{0pt}{3.5ex} \textbf{Dataset} & \textbf{Check-ins} & \textbf{Users} & \textbf{Locations} & \textbf{Links} & \textbf{Location Categories} \\ 
		
		\hline \rule[-2ex]{0pt}{3.5ex} Gowalla & 36,001,959 & 319,063 & 2,844,076 & 337,545 & 629 \\ 
		
		\hline \rule[-2ex]{0pt}{3.5ex} Weeplace & 7,658,368 & 15,799 & 971,309 & 59,970 & 96 \\
		
		\hline
	\end{tabular}
	\caption {Statistics of the datasets.} 
	\label{tab:dataset_stat} 
	%\vspace{-2em}
	\centering
\end{table}

After avoiding incomplete records, the 5 most checked-in categories (and their check-in count) were: (i) Home/Work/Other: Corporate/Office (437,824), (ii) Food: Coffee Shop (267,589), (iii) Nightlife:Bar (248,565), (iv) Shop: Food\& Drink:Grocery/Supermarket (161,016), and (v) Travel: Train Station (152,114) for Weeplaces, and (i) Corporate Office (1,750,707), (ii) Coffee Shop (1,063,961), (iii) Mall (958,285), (iv) Grocery (884,557), and (iv) Gas \& Automotive (863,199) for the Gowalla dataset.
The \textit{work} or \textit{home}-related category (Home/Work/ Other:Corporate/Office) was popular from 6 am to 6 pm, with the highest check-ins (42,019) made at 1 pm. Similarly, the bars had highest of 21,806 check-ins at 2 am and the lowest check-ins (15,209) at 5 am.
Most of the check-ins were at 12 pm - 6 pm and were either in Home or Work related categories.

\begin{figure}[h!]
	\centering
	\includegraphics[width=0.5\columnwidth,height=0.25\columnwidth]{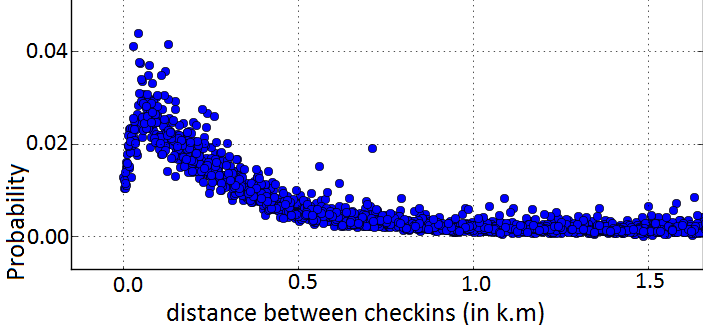}%
	\caption{Spatial impact on Check-ins}%
	\label{fig:checkins_distance_weeplace}
	%\vspace{-1.5em}
\end{figure}

\begin{figure}[h!]
	\centering     %%% not \center
	\subfigure[Check-ins of top user]{\label{fig:checkin_dist_top_1_users}\includegraphics[width=0.32\textwidth,height=0.13\textheight]{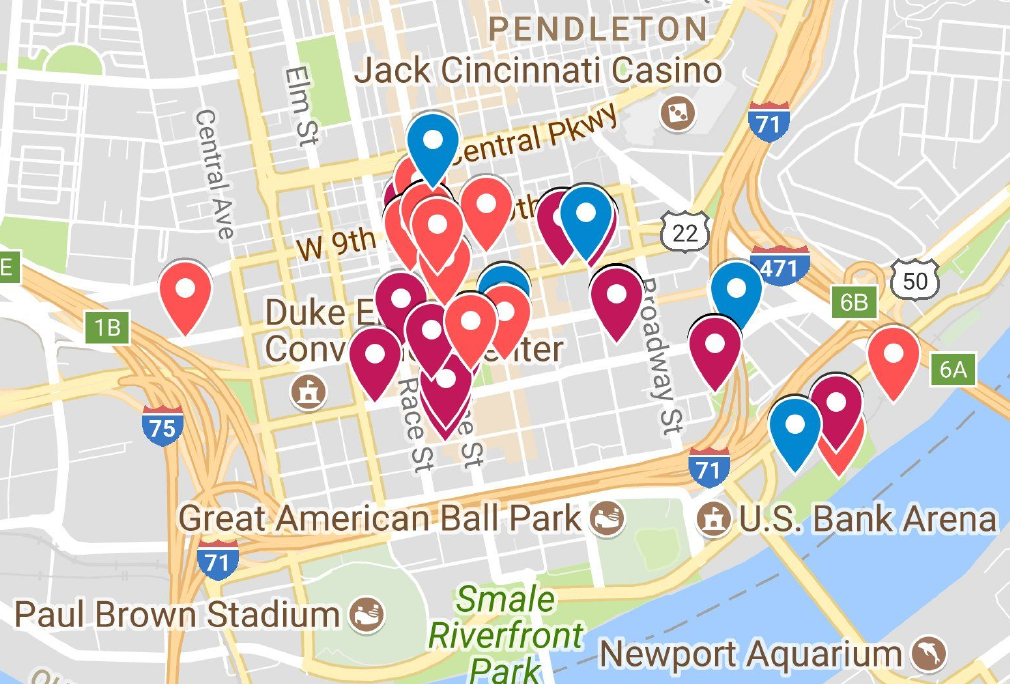}}
	\subfigure[Check-ins of $2^{nd}$ top user]{\label{fig:checkin_dist_top_2_users}\includegraphics[width=0.32\textwidth,height=0.13\textheight]{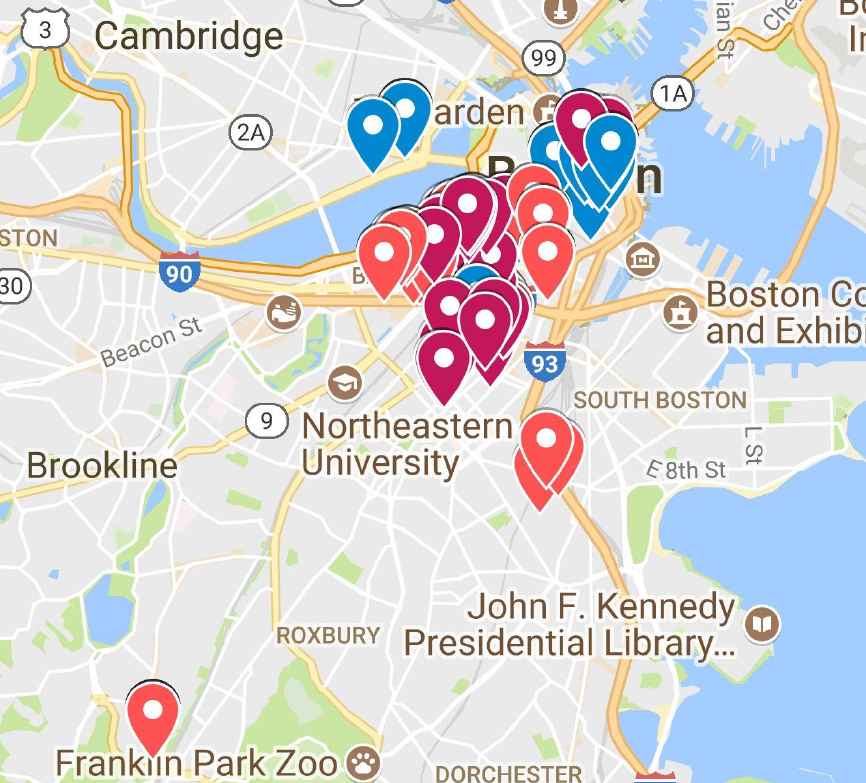}}
	\subfigure[Check-ins of $3^{rd}$ top user]{\label{fig:checkin_dist_top_3_users}\includegraphics[width=0.32\textwidth,height=0.13\textheight]{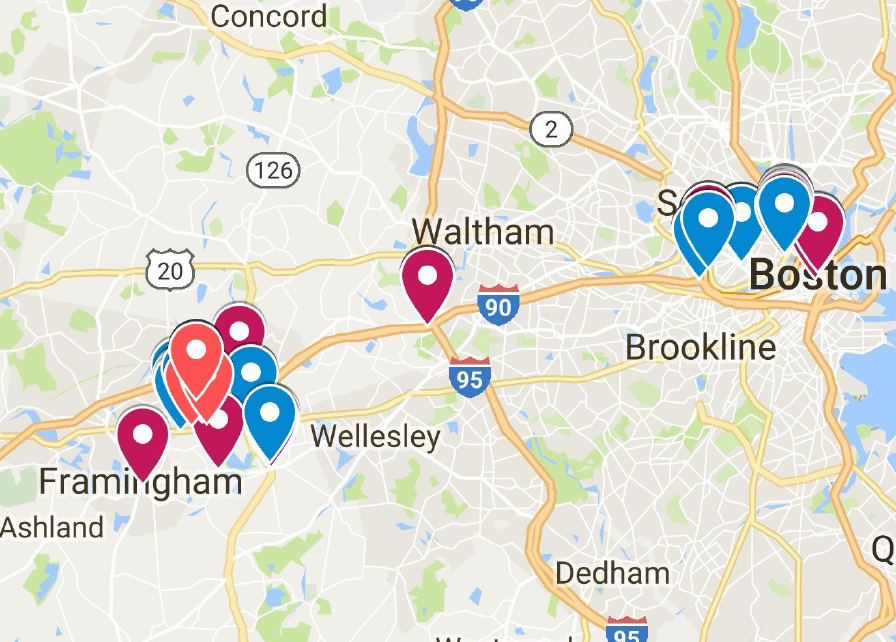}}
		\caption{Check-in distribution of 3 highest check-in days of top three users in Weeplaces dataset (Three different color marks are used for three different days. Repeated check-ins are overlapped and not distinguished. All three users have check-ins within reasonable distance and the sequences are of reasonable length.)}
	\label{fig:checkin_dist_top_users}
\end{figure}

The check-in distribution of all users in the Weeplaces and Gowalla dataset is illustrated in Figure~\ref{fig:checkin_dist}. Figure~\ref{fig:daily_checkins_weeplace} and Figure~\ref{fig:daily_checkins_gowalla} show that the frequency of daily check-ins is $\textless$50 for most of the users. This implies that most of the users have daily sequence length that is reasonable and exploiting the proposed model within this sequence length is enough to evaluate its performance.
Among the many factors influencing the check-in trend of the users, distance measure is one of the major factors. Figure~\ref{fig:checkins_distance_weeplace} illustrates the inverse relation between the distance of a location and likelihood of check-ins (i.e. most users preferred near places ($\leq$1 K.m.)). This implies that most of the users prefer to visit near places and hence the POIs within a sequence should also be within a reasonable distance.
The three days' check-in distribution of three users with most check-ins in Weeplaces dataset is illustrated in Figure~\ref{fig:checkin_dist_top_users}. Three different color marks are used to distinguish the check-in of different days. 
The overlapping check-ins on different days are overlapped in the map and are not distinguishable.
This figure also illustrates that most of the users prefer check-ins to near locations and most of the users have reasonably smaller sequence length.

\begin{figure*}[h!]
	\centering     
	\subfigure[Distribution of daily check-ins in Weeplaces dataset]{\label{fig:daily_checkins_weeplace}\includegraphics[width=0.45\textwidth,height=0.25\textheight]{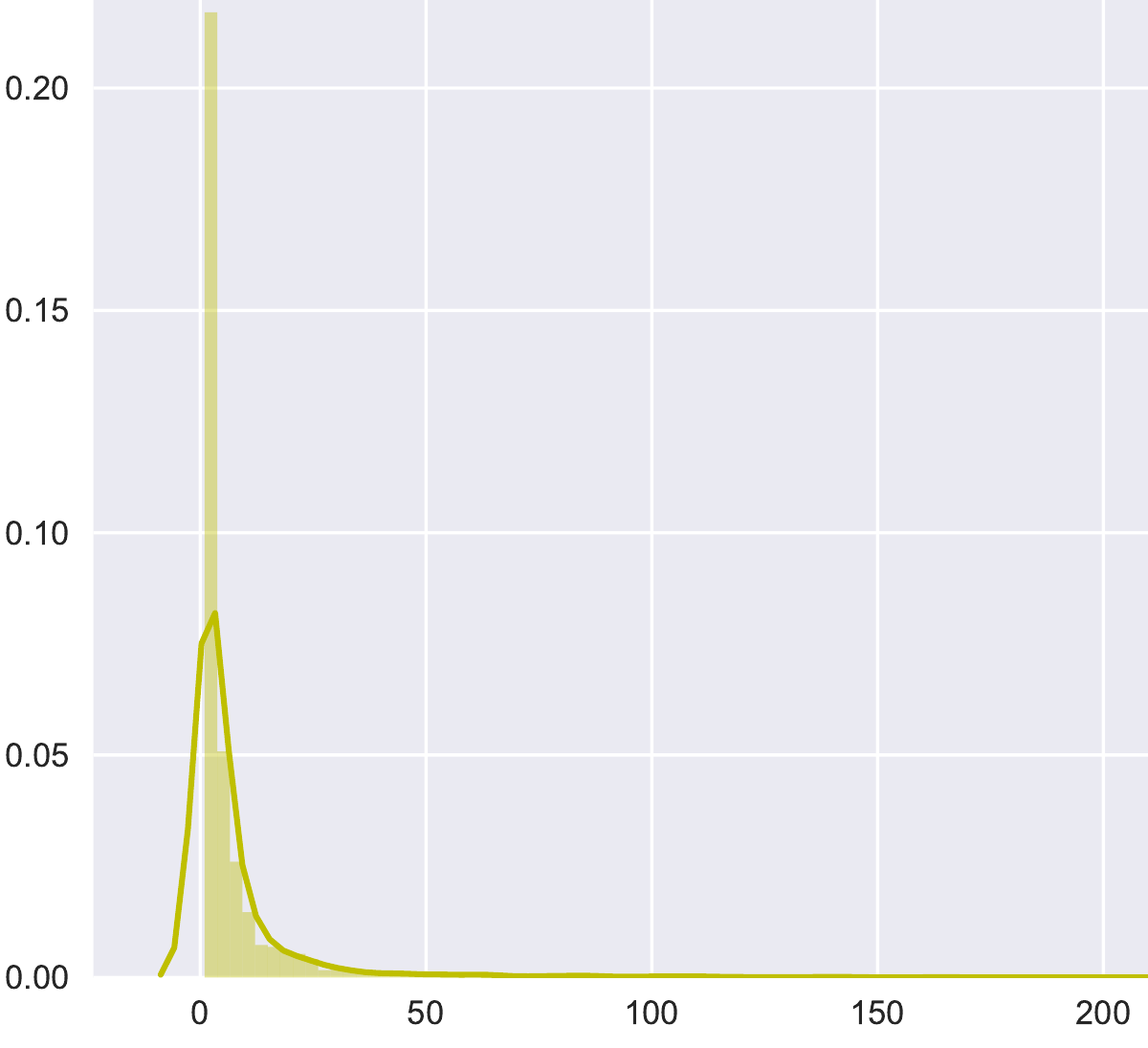}}
	\subfigure[Distribution of daily check-ins in Gowalla dataset]{\label{fig:daily_checkins_gowalla}\includegraphics[width=0.45\textwidth,height=0.25\textheight]{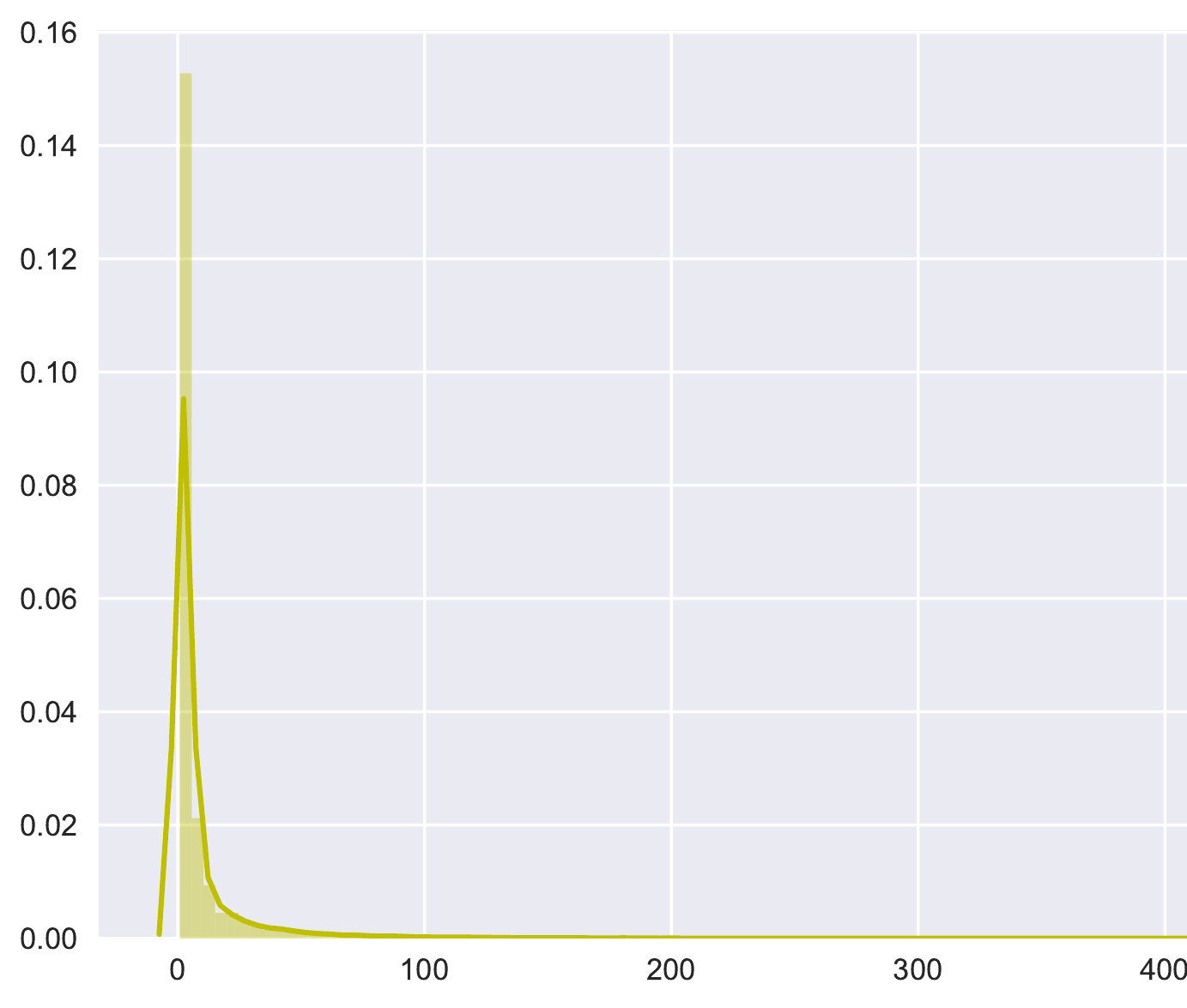}}
	\caption{(a) and (b) show the distribution of daily check-ins (X-axis represents the number of daily check-ins and Y-axis represents the fraction of users with the respective number of check-ins)}
	\label{fig:daily_checkins}
\end{figure*}

Unlike other studies, we did not split the dataset into different cities because we found that most of the users have check-ins spanned across multiple cities and splitting the dataset into multiple cities can disrupt the coherence of the places the users have in the dataset. The check-ins were chronologically sorted and for every user, the check-ins were splitted into subsequence of every 8 hours.

%\subsection{Hierarchical-approach for sequence generation}
%\label{ssec:hierarchical_seq}
%\include{hierarchical_sequence}

\subsection{Evaluation Metrics}
\label{ssec:eval_metrics}
It is difficult to measure the correctness of the order of items using simple precision, recall, and F-score. We use the pairs-F1 metrics defined in ~\cite{chen2016learning}. It considers both the POI identity and its order by using the F1 score of every pair of POIs in a sequence and is defined as:
\begin{equation}
pairs-F1 =  \frac{2* P_{PAIR} * R_{PAIR}}{P_{PAIR} + R_{PAIR}},
\end{equation}
where $P_{PAIR}$ and $R_{PAIR}$ are the precision and recall of the ordered POI pairs respectively.

We also evaluate the diversity (\cite{bradley2001improving}) of locations in the sequence. The diversity of locations is measured using their categorical similarity (i.e. Similarity =1 if two places are of same category and Similarity = 0 otherwise). We use the following relation to define the diversity of items in the list:
\begin{equation}
\text{Diversity }(c_1, c_2,...,c_n) = \frac{ \sum\limits_{i=1}^{n} \sum\limits_{j=i+1}^{n} (1 - \text{Similarity }(c_i, c_j))} {\frac{n}{2}*(n-1)}
\end{equation}

We use the displacement metric to measures the distance (in K.m.) between the predicted POIs and the actual POIs in the sequence. It is defined as:
\begin{equation}
\text{Displacement }(seq_{a}, seq_{e}) = \sum\limits_{i=1}^{k}\mid \text{Distance}(seq_{a_i}, seq_{e_i}) \mid,
\end{equation}
where $seq_a$ and $seq_e$ are the actual and estimated sequences respectively.

\subsection{Evaluation Baselines}
\label{ssec:eval_baseline}
We compared our proposed method with the following baseline models:

\begin{enumerate}
	\item POI Popularity: This is a naive approach that relies on the popularity of POIs. For a given location, an area within a predefined radius is used to find the most popular POI (i.e. most visits in the locality) within that area which is not already included in the list. The radius is dynamically updated by a predefined factor when no location is found in the area.
	
	\item 	Markov Chain-based approach: We use first order Markov Chain to generate the POI sequences. The Laplace smoothed state-transition matrix and initial probability matrix are derived from the check-in data and are personalized for each user.
	
	\item 	Apriori-based approach:  The most frequently checked-in place of a user is considered as the starting point because the model does not have provision of user inputs. From the starting point, we select the places that are within some threshold distance ($\epsilon$) to get the candidate 1-sets. The candidate 1-sets are used to get the other candidate sets. 
	All the candidates that do not satisfy the constraints are pruned. The constraint checking procedure and candidate generation procedure continues till the trip of desired length is obtained or the candidate sets are exhausted. Every trip that has greater than 8 hours travel time are pruned. Basically, we adapt a greedy pruning approach to get rid of the less preferred routes. Among the available candidate sets, we select the one that satisfies the following criteria: (i) has higher trip score, and (ii) has lower travel time (see Eqn~\ref{eqn:cons_score}).
	The trip score is calculated by adding the preference scores (see Eqn.~\ref{eqn:preference_score} and Eqn.~\ref{eqn:cons_score}) of all the places in the trip. The top-k trips with highest scores are recommended to the user. 
	Some of the existing studies (\cite{lu2012personalized, yu2016personalized}) have also exploited the Apriori-based approach.

	%\item Tree-based ~\cite{zhang2015personalized}
	
	%\item Ranking using RankSVM ~\cite{chen2016learning}
	
	\item	HITS-based approach (\cite{zheng2011learning}): In this, the locations are hierarchically organized into clusters/regions and the hub scores of user and authority scores of places are generated relevant to the regions. This facilitates in modeling the popularity of locations and users within a region. The inference is made by using the adjacent matrix between users and locations with respect to the region. The score of a sequence is determined using the hub scores of users who visited the sequence, and the authority scores of places weighted by the probability that people would consider the sequence. The top k popular sequences are recommended.
	
	%\item Yu et al.~\cite{yu2016personalized}
\end{enumerate}

%baseline approach to take k popular POIs.

%Show results when the individual checkin sequence is consider for one day, 8 hours, etc.

\subsection{Experimental settings}
We used a 5-fold cross validation to measure the performance of the models. An Ubuntu 14.04.5 LTS, 32 GB RAM, a Quadcore Intel(R) Core(TM) i7-3820 CPU @ 3.60 GHz was used to evaluate the models. The same configuration with a Tesla K20c 6 GB GPU was used to evaluate the neural network based models. 
%We used Networkx~\cite{hagberg-2008-networkx}, Numpy~\cite{walt2011numpy}, Pandas~\cite{mckinney-proc-scipy-pandas}, and TensorFlow~\cite{tensorflow2015-whitepaper} as the supplement programming libraries.
We used Python~\footnote{https://www.python.org} as the programming platform, Numpy~\footnote{http://www.numpy.org} as the mathematical computation library, Pandas~\footnote{http://pandas.pydata.org} as data analysis library, and TensorFlow~\footnote{https://www.tensorflow.org} as the neural network library.

%The daily check-in frequency of $\geq$7 was considered a valid sequence and the users with less than 20 valid sequences were ignored.
The users with less than 25 check-ins were ignored. 
The context and feature vector were estimated from the historical check-ins of the users. 
For each user, 7 most frequently checked-in places were taken as starting point, and 10 sequence per starting point was generated. The average metrics on these 10 sequences was observed. The POI-Popularity and Apriori models used distance threshold of 2 K.m.
The contextual LSTM used 512 hidden states, and the contextual RNN used 5 layers and 256 nodes. The input sequence length was set to 25, the data was fed in batches of size 50, embedding vectors were of size 384, and the experiment was repeated for 100 epochs.
The learning rate was set to 0.002, and the gradients were clipped at 5 to prevent overfitting.

\subsection{Experimental Results and Discussions}
The pair-wise precision, recall, and F-score of different models is illustrated in Table~\ref{tab:result_pair_score_weeplace} and Table~\ref{tab:result_pair_score_gowalla}. The diversity and displacement-based performance is illustrated in Table~\ref{tab:result_div_disp_weeplace} and Table~\ref{tab:result_div_disp_gowalla}.

\begin{table}[h!]
	\centering
	\begin{tabular}{|c|c|c|c|}
		\hline \textbf{Models} & \textbf{$Precision_{PAIR}$}& \textbf{$Recall_{PAIR}$} & \textbf{$Pair-F1$} \\
		\hline \textbf{Weeplaces Dataset}\\
		
		\hline POI-Popularity & 0.30000  & 0.16666  & 0.21428 \\

		\hline Apriori & 0.46079  & 0.23088  & 0.30762   \\
		
		\hline POI-Markov  & 0.49411  & 0.24711  & 0.32945 \\
		
		%\hline RankSVM  & 0.00000  & 0.00000  & 0.00000  \\
		
		%\hline Tree-based & 0.00000  & 0.00000  & 0.00000  \\
		
		\hline HITS & 0.49981  & 0.27336  & 0.35342   \\
		
		\hline Vanilla RNN & 0.49788  & 0.27618  & 0.35528  \\
		
		\hline LSTM & 0.51557  & 0.27500  & 0.35868  \\
		
		\hline CAPS-RNN & 0.62422  & 0.41970  & 0.50192  \\
		
		\hline CAPS-LSTM & 0.67771  & 0.43100  & \textbf{0.52690}$^*$  \\
		\hline
	\end{tabular}
	\caption {Pair F-Score Performance of different models ($^*$ implies observed difference was statistically significant
		at 95\% confidence level) on Weeplace dataset} \label{tab:title}
	\label{tab:result_pair_score_weeplace}
\end{table}

\begin{table}[h!]
	\centering
	\begin{tabular}{|c|c|c|c|}
		\hline \textbf{Models} & \textbf{$Precision_{PAIR}$}& \textbf{$Recall_{PAIR}$} & \textbf{$Pair-F1$} \\			
		\hline \textbf{Gowalla Dataset}\\
		
		\hline POI-Popularity & 0.36442  & 0.20010  & 0.25834 \\

		\hline Apriori & 0.46922  & 0.24276 & 0.31997   \\
		
		\hline POI-Markov  & 0.49993  & 0.24981  & 0.33314 \\
		
		%\hline RankSVM  & 0.00000  & 0.00000  & 0.00000  \\
		
		%\hline Tree-based & 0.00000  & 0.00000  & 0.00000  \\
		
		\hline HITS & 0.50653  & 0.27993  & 0.36058   \\
		
		\hline Vanilla RNN & 0.51001  & 0.27896  & 0.36065  \\
		
		\hline LSTM & 0.53333  & 0.44000  & 0.48219  \\
		
		\hline CAPS-RNN & 0.60914 & 0.43000  & 0.50412  \\
		
		\hline CAPS-LSTM & 0.67112  & 0.44462  & \textbf{0.53487}$^*$  \\
		\hline
	\end{tabular}
	\caption {Pair F-Score Performance of different models ($^*$ implies observed difference was statistically significant
		at 95\% confidence level) in Gowalla dataset} \label{tab:title}
	\label{tab:result_pair_score_gowalla}
\end{table}

\begin{table}[h!]
	\centering
	\begin{tabular}{|c|c|c|}
		\hline \textbf{Models} & \textbf{Diversity}& \textbf{Displacement(K.m.)} \\
		\hline \textbf{Weeplaces Dataset}\\
		
		\hline POI-Popularity & 1.20000  & 23.30785   \\

		\hline Apriori & 1.90000  & 13.00000  \\
		
		\hline POI-Markov  & 2.50000  & 11.72130  \\
		
		%\hline RankSVM  & 0.00000  & 0.00000  & 0.00000  \\
		
		%\hline Tree-based & 0.00000  & 0.00000  & 0.00000  \\
		
		\hline HITS & 4.00000  & 10.55233  \\
		
		\hline Vanilla RNN & 6.22000  & 10.27620 \\
		
		\hline LSTM & 6.73000  & 10.00023   \\
		
		\hline CAPS-RNN & \textbf{7.11820}  & 8.22990  \\
		
		\hline CAPS-LSTM & 7.09120  & \textbf{7.77014}  \\
		
		\hline
	\end{tabular}
	\caption {Diversity and displacement performance (on sequence length of 25) of models in Weeplace dataset} \label{tab:title}
	\label{tab:result_div_disp_weeplace}
	%\vspace{-2.5 em}
\end{table}

\begin{table}[h!]
	\centering
	\begin{tabular}{|c|c|c|}
		\hline \textbf{Models} & \textbf{Diversity}& \textbf{Displacement(K.m.)} \\
		\hline \textbf{Gowalla Dataset}\\
		
		\hline POI-Popularity & 3.20000  & 25.22877   \\

		\hline Apriori & 3.33500  & 13.00000  \\
		
		\hline POI-Markov  & 3.50000  & 11.22113  \\
		
		%\hline RankSVM  & 0.00000  & 0.00000  & 0.00000  \\
		
		%\hline Tree-based & 0.00000  & 0.00000  & 0.00000  \\
		
		\hline HITS & 4.00000  & 11.11224   \\
		
		\hline Vanilla RNN & 5.88441  & 11.00111   \\
		
		\hline LSTM & 7.22533  & 10.33333   \\
		
		\hline CAPS-RNN & 8.44765  & 7.77669  \\
		
		\hline CAPS-LSTM & \textbf{8.45001}  & \textbf{7.71001}  \\
		
		\hline
	\end{tabular}
	\caption {Diversity and displacement performance (on sequence length of 25) of models in Gowalla dataset} \label{tab:title}
	\label{tab:result_div_disp_gowalla}
	%\vspace{-2.5 em}
\end{table}

The popularity-based model performed worst among all the models. It generated almost similar sequences for all the users and hence was not relevant to personalized preferences. This might be due to the ignorance of the personalized preferences of user. The diversity measure was also quite low, which means the POIs in the generated sequences included few categories. The high displacement metrics indicates that the predicted POIs were far from the actual ones.

The Apriori-based model had better performance than popularity-based model. Although the Apriori-based model pruned irrelevant candidate sequences, it also did not capture the personalization aspect. This might be the reason behind its low performance. The diversity and displacement measures were also better than the popularity-based model.

The first-order Markov model relied on one previous check-in data to determine next location and hence was not able to fully model the check-in sequence generation process. However, its pair-Fscore, diversity and displacement metrics were better than popularity-based and Apriori-based models which is due to the personalization implied from separate initial-probability and state-transition tables for each user.

The HITS-based model slightly outperformed Markov-based model on all three metrics. As it relies on segregation of places into regions and finding the authority and hub scores of the places and users within that regions, its performance depends on the region generation approach.
We used a radius of 10 K.m. from a specified location to generate such regions. Its performance with the radius of 5 K.m. and 15 K.m. was in par with popularity-based model.

Though not sophisticated, the performance of vanilla RNN was in par with the HITS model. This might be because of its capability to retain information of previous items from the sequence. The regular LSTM performed slightly better than vanilla RNN due to its implicit ability to cope with vanishing gradient problem.

The \textbf{CAPS-LSTM} model was best performer in both dataset. Its performance was slightly better than CAPS-RNN in all the cases except the diversity metrics in Weeplaces dataset where CAPS-RNN was slightly better. The performance of CAPS-LSTM was significant in larger dataset (i.e. Gowalla).
This is obvious because the Gowalla dataset is larger, and has more social (friendship) relations, which favored the neural networks that need lots of training data for better performance. We used the threshold of 6, 8, and 10 hours and found that the threshold of 8 hours performed better than others. This might however, vary on the dataset used.
In terms of execution time, \textbf{CAPS-LSTM} was slower as it used multiple epochs and required lots of training time. The execution time of different models were in the order of: CAPS-LSTM $\geq$ CAPS-RNN $\textgreater$ LSTM $\ge$ vanilla RNN $\textgreater$ HITS $\textgreater$ Apriori $\ge$ Markov $\textgreater$ Popularity-based.

To summarize, the CAPS-LSTM performed best, followed by the CAPS-RNN in all three metrics. 
This evaluation result supports our claim that extension of RNN and it's variants by incorporating item-wise contexts and sequence-wise features can be efficient for sequence modeling.

\begin{figure*}[h!]
	\centering     
	\subfigure[Diversity performance on Weeplace dataset]{\label{fig:daily_checkins_weeplace}\includegraphics[width=0.45\textwidth,height=0.35\textheight]{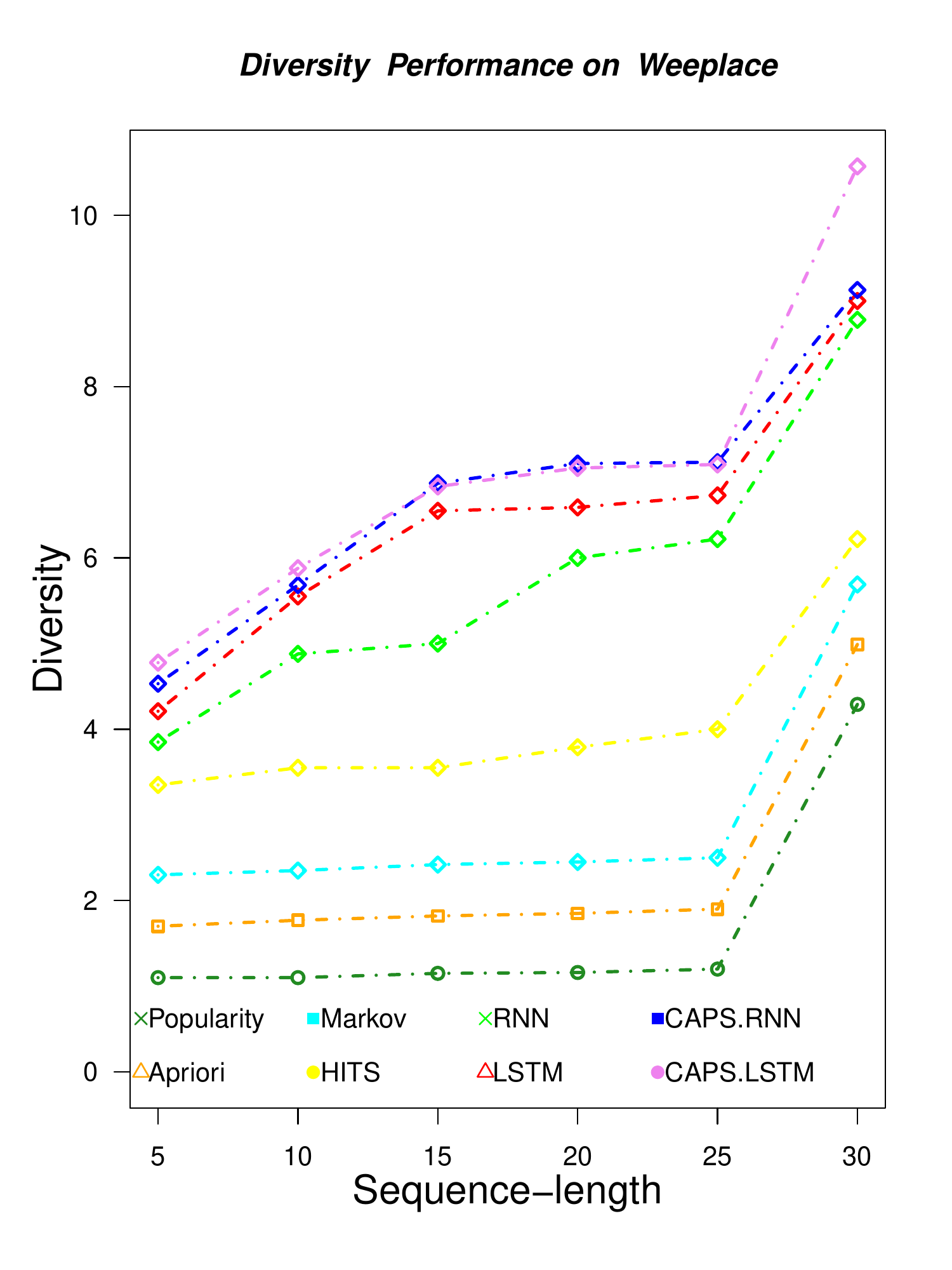}}
	\subfigure[Diversity performance on Gowalla dataset]{\label{fig:daily_checkins_gowalla}\includegraphics[width=0.45\textwidth,height=0.35\textheight]{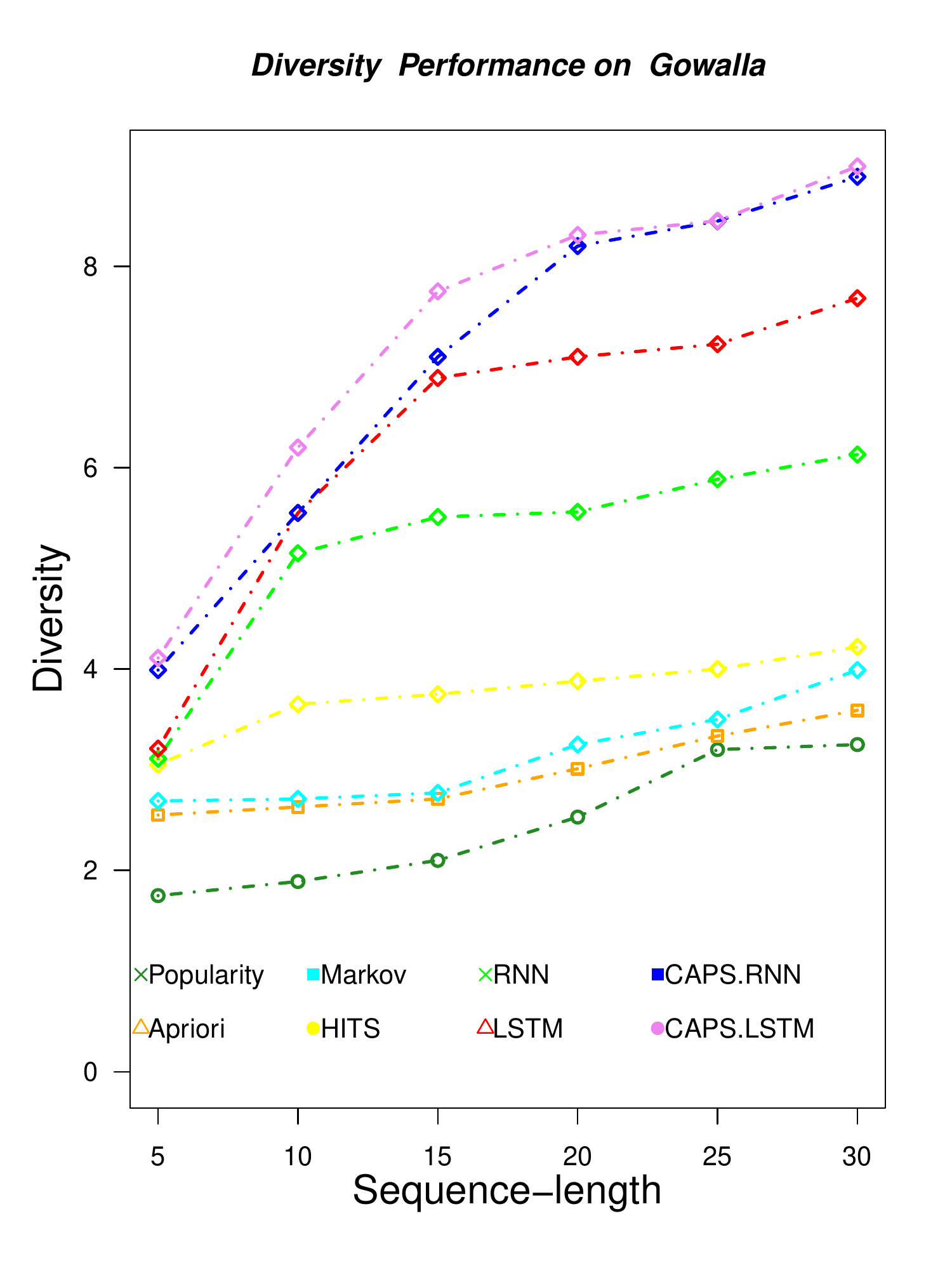}}
	\caption{(a) and (b) show the trend in diversity with the increasing sequence length}
	\label{fig:diversity_trend}
	%\vspace{-1.5em}
\end{figure*}

\begin{figure*}[h!]
	\centering     
	\subfigure[Displacement performance on Weeplace dataset]{\label{fig:daily_checkins_weeplace}\includegraphics[width=0.45\textwidth,height=0.35\textheight]{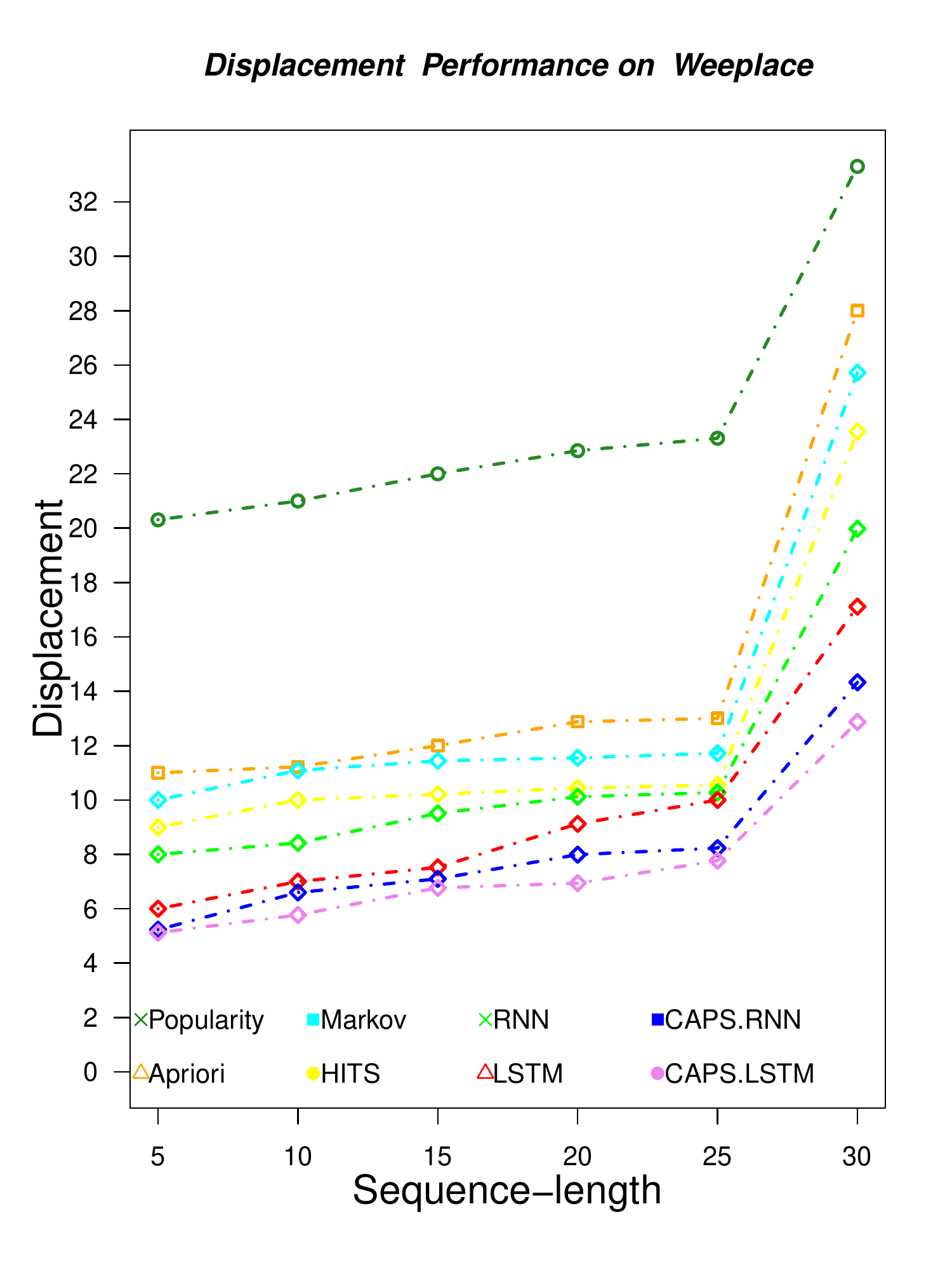}}
	\subfigure[Displacement performance on Gowalla dataset]{\label{fig:daily_checkins_gowalla}\includegraphics[width=0.45\textwidth,height=0.35\textheight]{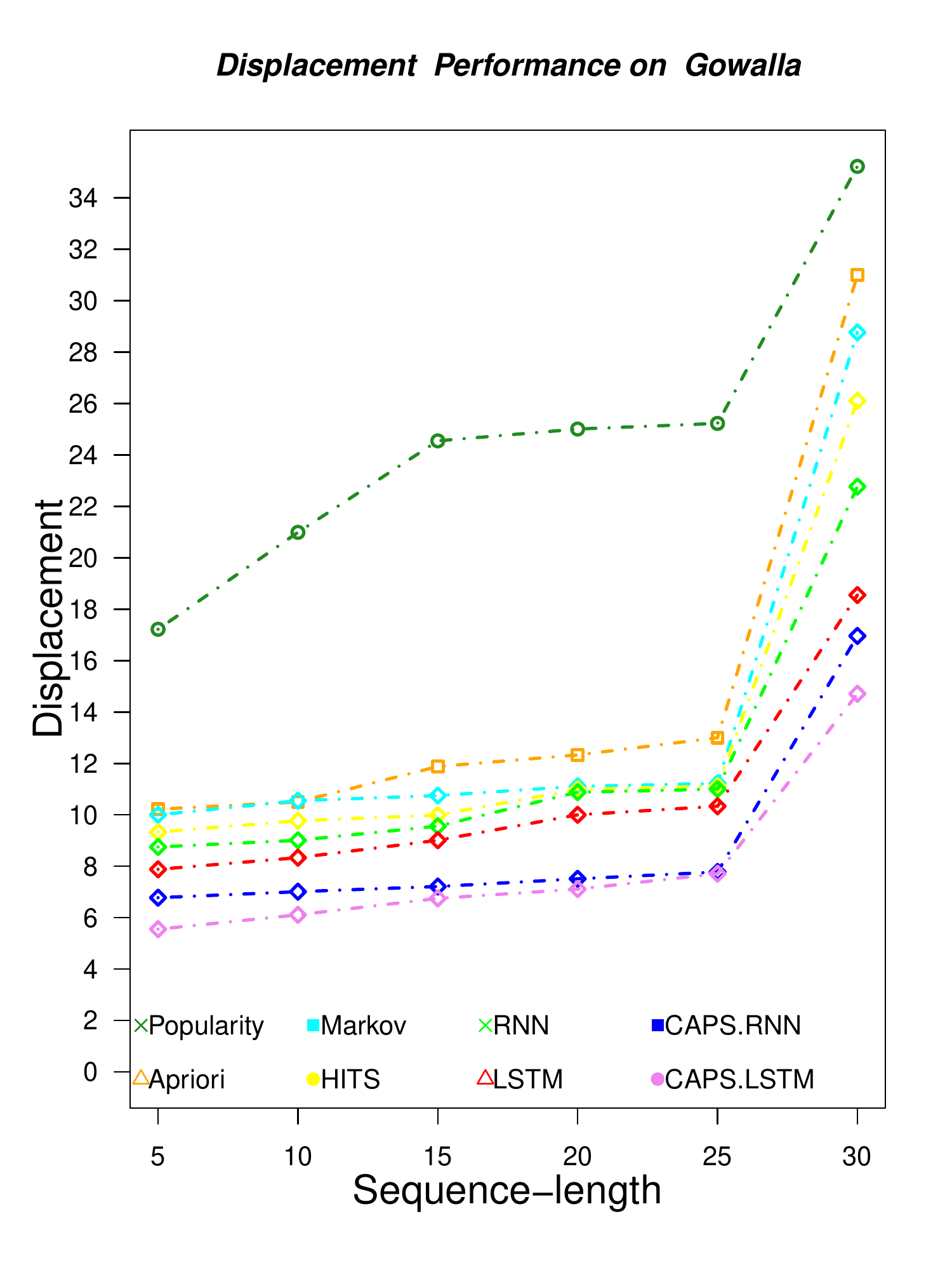}}
	\caption{(a) and (b) show the trend in displacement with the increasing sequence length}
	\label{fig:displacement_trend}
	%\vspace{-1.5em}
\end{figure*}

\subsection{Impact of sequence length}

Figure~\ref{fig:diversity_trend} illustrates the trend of diversity changes with the increasing sequence length. 
The diversity metric measures the extent to which different categories are included in the recommendation. A good recommendation should include items with different category that satisfy the user preferences. We can see that the CAPS-RNN and CAPS-LSTM have better diversity trend with the increasing sequence length. It is most likely to have locations of different category when sequence length is increased. This effect is clearly reflected in all neural network-based models in both datasets. The popularity model always recommends popular items and does not focus on the diversity, hence its diversity is lowest of all the models. Although the Apriori, Markov, and HITS models perform better than popularity-based model, their performance is lower than that of the neural network-based models. The increase in diversity was quite slow as we increase the sequence length. This might be because of the fine grained categories which cover only few places and most of the broader categories cover many items and after some sequence length, the newly added places overlap with the category of places already in the sequence.

Figure~\ref{fig:displacement_trend} illustrates the impact of sequence length on the displacement. As the popularity model recommends only the popular items no matter how far they are, it has highest displacement among all the models. The displacement is increasing with the increasing sequence length. The CAPS-RNN and CAPS-LSTM show similar effect of sequence length on the displacement. The other models have reasonable displacement until we reach the sequence of length 25.
For all the models, the increase in displacement is sharp after sequence length of 25. This is also supported by the check-in trend presented in Figure~\ref{fig:daily_checkins} which shows that the daily check-in frequency of a user is $\sim$25 and as we increase the sequence length, the check-ins are most likely to be made on another day (which might be in another city or some place farther than the one checked-in on previous days). As the CAPS-RNN and CAPS-LSTM have better performance for subsequences, they have better performance over other models.

\section{Case study on generated trajectories}
\begin{comment}
\textcolor{red}{In the results section, it would be nice to have a more in-depth characterization of the trajectories generated. How long are they? How do the performance metrics vary with their length? Section 3.1.5 could give some additional detail on the generation process and the evaluation procedure (maybe adding some examples?).
\newline
Take case of two users, where we have a sequence of length 5 from popularity model, one from CAPS-LSTM, and one from RNN for both users. 
Using real data show how the performance (diversity and displacement) vary on the two users when the sequence is generated using these three models. This should explain that the sequence is taken from the recommended sequence and how the CAPS-LSTM was better in recommending sequence of reasonable diversity and displacement measure.}
\end{comment}
In this section, we provide a case study on POI sequences generated by popularity-based approach and CAPS-LSTM on Gowalla dataset. We select sequences of length 5 for two different users 'thadd-fiala', 'boon-yap' (known as $u_1, u_2$ in rest of the paper) with most check-ins and analyze the relevance of the sequences for them.
For $u_1$, a sequence of length 5 from popularity model is \textbraceleft'sycamore-place-lofts-cincinnati', 'pg-gardens-cincinnati', 'lytle-park-cincinnati', 'piatt-park-cincinnati', 'sycamore-place-at-st-xavier-park-apartments-cincin'\textbraceright, and their respective categories are \textbraceleft'Home/Work/Other:Home', 'Parks \& Outdoors:Plaza / Square', 'Parks \& Outdoors:Park', 'Parks \& Outdoors:Plaza / Square', 'Home/Work/Other:Home'\textbraceright. 
Similarly for user $u_2$ a length 5 sequence is \textbraceleft'starbucks-boston', 'mbta-south-station-boston', 'boston-common-boston', 'dunkin-donuts-boston', 'mbta-park-street-station-boston'\textbraceright and their respective categories are \textbraceleft'Food:Coffee Shop', 'Travel:Train Station', 'Parks \& Outdoors:Park', 'Food:Donuts', 'Travel:Train Station'\textbraceright.
Most of the places recommended are the popular ones and the generated sequences have less diversity. For both users, there are three different categories in the generated sequences. With the increasing sequence length, the diversity shows some increasing trend (see Figure~\ref{fig:diversity_trend}) but this is lower in both datasets.

With \textbf{CAPS-LSTM}, a sequence generated for user $u_1$ is \textbraceleft'sycamore-place-lofts-cincinnati', 'pg-gardens-cincinnati', 'piatt-park-cincinnati', 'lytle-park-cincinnati','lpk-cincinnati'\textbraceright and their categories are \textbraceleft'Home/Work/ Other: Home', 'Parks \& Outdoors:Plaza/Square', 'Parks \& Outdoors:Plaza/Square', 'Parks \& Outdoors:Park', 'Home/Work/ Other:  Corporate/ Office'\textbraceright. 
For user $u_2$ a sequence is \textbraceleft'starbucks-boston', 'mbta-park-street-station-boston', 'boston-common-boston', 'digitas-boston-boston', 'hubspot-cambridge'\textbraceright and their categories are \textbraceleft'Food:Coffee Shop', 'Travel:Train Station', 'Parks \& Outdoors: Park', 'Nightlife: Speakeasy / Secret Spot', 'Home/Work/Other: Corporate/ Office'\textbraceright. We can observe that for both users, there are four different categories in the sequence and the recommendation is more contextual.
With the increasing sequence length, the diversity shows some increasing trend (see Figure~\ref{fig:diversity_trend}) which is better with \textbf{CAPS-LSTM} among all other models.

With the popularity-based model, the average displacement of above sequence was 19.36 K.M. for user $u_1$ and it was 20.03 K.M. for user $u_2$. With the CAPS-LSTM, the average displacement of above sequence was 5.02 K.M for user $u_1$ and it was 5.61 K.M. for user $u_2$. This shows that CAPS-LSTM addresses the distance constraint better for both users.
With the increasing sequence length, the displacement trend increased for both models and followed the trend as shown in Figure~\ref{fig:displacement_trend}.

\paragraph{\textbf{Limitations}}
%\vspace{-0.5em}
The deep models need lots of training data and the valid check-in sequences (the sequences that have minimum number of check-ins) that we used might still be insufficient to exploit the full potential of the proposed model.
As the model needs to be trained offline, the model in its current state might not be efficient for real-time prediction. 
The threshold of 8 hours might not be equally applicable for all users, all types of trips, and might vary across different datasets.

\section{Conclusion and Future Work}
We formulated the contextual personalized POI sequence modeling problem by extending the recurrent neural networks and its variants. We incorporated different contexts (such as social, temporal, categorical, and spatial) by feeding them to the hidden layer and the output layer. We propagated the feature vector to all the layers of the network and retained the contextual information that was valid throughout the sequence. 
We evaluated the proposed model with two real-world datasets and demonstrated that the contextual models can perform better than regular models.
The proposed model performed slightly better with larger datasets. There are many interesting directions to explore. We would like to incorporate the textual attributes (e.g., tags, tips, and review text) and also visual information (e.g., image of places) to define the preference of users and popularity of places. We would also like to explore other datasets for sequence modeling.

%\begin{acknowledgements}
%\end{acknowledgements}
\bibliographystyle{aps-nameyear}      % American Physical Society (APS) style, author-year citations
\bibliography{poi_sequence}                % name your BibTeX data base
%\nocite{*}

\end{document}